\documentclass[authoryear,preprint,review,12pt]{elsarticle}

\usepackage{amssymb}
\usepackage[bookmarks = true, bookmarksnumbered = true, pdfpagemode =None, pdfstartview = FitH, pdfpagelayout = SinglePage, colorlinks = true, urlcolor = blue, citecolor = monbleu]{hyperref}
\usepackage{threeparttable}

\newcommand{\Mearth}{$M_\oplus$}

\newcommand{\Kunit}{\,cm$^{2}\cdot$s$^{-1}$}
\newcommand{\dix}[1]{$\times10^{#1}$}
\newcommand{\fig}[1]{Fig.~\ref{#1}}
\newcommand{\pderiv}[2]{\frac{\partial#1}{\partial#2}}

\usepackage{lineno}

\journal{Planetary and Space Science}

\begin{document}

\begin{frontmatter}



\title{Scientific rationale of Saturn's {\it in situ} exploration}

\author[UTI]{O. Mousis}
\ead{olivier.mousis@obs-besancon.fr}
\address[UTI]{Universit{\'e} de Franche-Comt{\'e}, Institut UTINAM, CNRS/INSU, UMR 6213, Observatoire des Sciences de l'Univers de Besan\c con, France\fnref{label2}}
\author[OX]{L. N. Fletcher}
\address[OX]{Atmospheric, Oceanic \& Planetary Physics, Department of Physics, University of Oxford, Clarendon Laboratory, Parks Road, Oxford OX1 3PU, UK}
\author[LPC2E,LESIA]{J.-P. Lebreton}
\address[LPC2E]{LPC2E, CNRS-Universit{\'e} d'Orl{\'e}ans, 3a Avenue de la Recherche Scientifique, 45071 Orl{\'e}ans Cedex 2, France}
\address[LESIA]{LESIA, Observatoire de Paris, CNRS, UPMC, Univ. Paris-Diderot, 5, place Jules Janssen, F-92195 Meudon Cedex}
\author[UBE]{P. Wurz}
\address[UBE]{Space Science \& Planetology, Physics Institute, University of Bern, Sidlerstrasse 5, 3012 Bern, Switzerland}
\author[MP]{T. Cavali{\'e}}
\address[MP]{Max-Planck-Institut f\"ur Sonnensystemforschung, Max-Planck-Str. 2, 37191 Katlenburg-Lindau, Germany}
\author[LESIA]{A. Coustenis}
\author[LESIA]{R. Courtin}
\author[LESIA]{D. Gautier}
\author[IS]{R. Helled}
\address[IS]{Department of Geophysics, Atmospheric and Planetary Sciences, Tel-Aviv University, Tel-Aviv, Israel}
\author[OX]{P. G. J. Irwin}
\author[OPEN]{A. D. Morse}
\address[OPEN]{Planetary and Space Sciences, Department of Physics, The Open University, Walton Hall, Milton Keynes MK7 6AA, UK}
\author[DE]{N. Nettelmann}
\address[DE]{Institute for Physics, University of Rostock, 18051 Rostock, Germany}
\author[CRPG]{B. Marty}
\address[CRPG]{CRPG-CNRS, Nancy-Universit\'e, 15 rue Notre Dame des Pauvres, 54501 Vandoeuvre-ls-Nancy, France}
\author[UTI]{P. Rousselot}
\author[BE]{O. Venot}
\address[BE]{Instituut voor Sterrenkunde, Katholieke Universiteit Leuven, Celestijnenlaan 200D, 3001 Leuven, Belgium}
\author[OH,JPL]{D. H. Atkinson}
\address[OH]{Department of Electrical and Computer Engineering, University of Idaho, Moscow ID 83844-1023, USA}
\author[SWRI]{J. H. Waite}
\address[SWRI]{Southwest Research Institute (SwRI), 6220 Culebra Road, San Antonio, TX 78228, USA}
\author[JPL]{K. R. Reh}
\address[JPL]{Jet Propulsion Laboratory, California Institute of Technology, 4800 Oak Grove Dr., Pasadena, CA 91109, USA}
\author[GD]{A. Simon-Miller}
\address[GD]{NASA Goddard Space Flight Center, Code 690, Greenbelt, MD 20771, USA}
\author[UMI]{S. Atreya}
\address[UMI]{Department of Atmospheric, Oceanic, and Space Sciences, University of Michigan, Ann Arbor, MI 48109-2143, USA}
\author[IRAP]{N. Andr\'e}
\address[IRAP]{ Institut de Recherche en Astrophysique et Plan\'etologie (IRAP), CNRS/Universit\'e Toulouse III (UMR 5277), 9, avenue du Colonel Roche, BP 44346, 31028 Toulouse Cedex 4, France, France}
\author[IRAP]{M. Blanc}
\author[GR]{I. A. Daglis}
\address[GR]{University of Athens, Department of Physics, Panepistimioupoli Zografou, 15784 Athens, Greece}
\author[AT]{G. Fischer}
\address[AT]{Space Research Institute, Austrian Academy of Sciences, Schmiedlstrasse 6, A-8042 Graz, Austria}
\author[SW]{W. D. Geppert}
\address[SW]{Stockholm University Astrobiology Centre, Department of Physics, AlbaNova, Stockholm University/Stockholms universitet, Roslagstullbacken 21, S-10691 Stockholm, Sweden/Sverige}
\author[OCA]{T. Guillot}
\address[OCA]{Observatoire de la C\^ote d'Azur, Laboratoire Lagrange, BP 4229, 06304 Nice cedex 4, France}
\author[IN]{M. M. Hedman}
\address[IN]{Department of Astronomy, Indiana University, Bloomington, Indiana 47405, USA}
\author[ES]{R. Hueso}
\address[ES]{Departamento F\' isica Aplicada I, Universidad del Pa\'is Vasco UPV/EHU, ETS Ingenier\'ia, Alameda Urquijo s/n, 48013 Bilbao, Spain\\
Unidad Asociada Grupo Ciencias Planetarias UPV/EHU-IAA(CSIC), 48013 Bilbao, Spain}
\author[LESIA]{E. Lellouch}
\author[CN]{J. I. Lunine}
\address[CN]{Center for Radiophysics and Space Research, Space Sciences Building, Cornell University, Ithaca, NY 14853, USA}
\author[UL]{C. D. Murray}
\address[UL]{School of Physics and Astronomy, Queen Mary University of London, Mile End Road, London E1 4NS, UK}
\author[AUL]{J. O'Donoghue}
\address[AUL]{Department of Physics and Astronomy, University of Leicester, Leicester LE1 7RH, UK}
\author[MP]{M. Rengel}
\author[ES]{A. S\'anchez-Lavega}
\author[OCA]{F.-X. Schmider}
\author[LMD]{A. Spiga}
\address[LMD]{Laboratoire de M\'et\'eorologie Dynamique, Universit\'e Pierre et Marie Curie, Institut Pierre Simon Laplace, Paris, France}
\author[SSSE]{T. Spilker}
\address[SSSE]{Solar System Science \& Exploration, Monrovia, USA}
\author[UTI]{J.-M. Petit}
\author[CN]{M. S. Tiscareno}
\author[UTI]{M. Ali-Dib}
\author[UBE]{K. Altwegg}
\author[UTI,SWRI]{A. Bouquet}
\author[LPC2E]{C. Briois}
\author[LESIA]{T. Fouchet}
\author[LMD]{S. Guerlet}
\author[GD]{T. Kostiuk}
\author[ALC]{D. Lebleu}
\address[ALC]{Thales Alenia Space, Cannes, France}
\author[LESIA]{R. Moreno}
\author[JPL]{G. S. Orton}
\author[ALC]{J. Poncy}

\begin{abstract}

Remote sensing observations meet some limitations when used to study the bulk atmospheric composition of the giant planets of our solar system. A remarkable example of the superiority of {\it in situ} probe measurements is illustrated by the exploration of Jupiter, where key measurements such as the determination of the noble gases abundances and the precise measurement of the helium mixing ratio have only been made available through {\it in situ} measurements by the Galileo probe. This paper describes the main scientific goals to be addressed by the future {\it in situ} exploration of Saturn placing the Galileo probe exploration of Jupiter in a broader context and before the future probe exploration of the more remote ice giants. {\it In situ} exploration of Saturn's atmosphere addresses two broad themes that are discussed throughout this paper: first, the formation history of our solar system and second, the processes at play in planetary atmospheres. In this context, we detail the reasons why measurements of Saturn's bulk elemental and isotopic composition would place important constraints on the volatile reservoirs in the protosolar nebula. We also show that the {\it in situ} measurement of CO (or any other disequilibrium species that is depleted by reaction with water) in Saturn's upper troposphere would constrain its bulk O/H ratio. We compare predictions of Jupiter and Saturn's bulk compositions from different formation scenarios, and highlight the key measurements required to distinguish competing theories to shed light on giant planet formation as a common process in planetary systems with potential applications to most extrasolar systems. {\it In situ} measurements of Saturn's stratospheric and tropospheric dynamics, chemistry and cloud-forming processes will provide access to phenomena unreachable to remote sensing studies. Different mission architectures are envisaged, which would benefit from strong international collaborations, all based on an entry probe that would descend through Saturn's stratosphere and troposphere under parachute down to a minimum of 10 bars of atmospheric pressure. We finally discuss the science payload required on a Saturn probe to match the measurement requirements.\\

\end{abstract}


\end{frontmatter}


\section{Introduction}
\label{intro}

Giant planets contain most of the mass and the angular momentum of our planetary system and must have played a significant role in shaping its large scale architecture and evolution, including that of the smaller, inner worlds \citep{2005Natur.435..466G}. Furthermore, the formation of the giant planets affected the timing and efficiency of volatile delivery to the Earth and other terrestrial planets \citep{2001M&PS...36..381C}. Therefore, understanding giant planet formation is essential for understanding the origin and evolution of the Earth and other potentially-habitable environments throughout our solar system. The origin of the giant planets, their influence on planetary system architectures, and the plethora of physical and chemical processes at work within their atmospheres, make them crucial destinations for future exploration. Because Jupiter and Saturn have massive envelopes essentially composed of hydrogen and helium and (possibly) a relatively small core, they are called gas giants. Meanwhile, Uranus and Neptune also contain hydrogen and helium atmospheres but, unlike Jupiter and Saturn, their H$_2$ and He mass fractions are smaller (5 to 20\%). They are called ice giants because their density is consistent with the presence of a significant fraction of ices/rocks in their interiors. Despite this apparent grouping into two classes of giant planets, the four giant planets likely exist on a continuum, each a product of the particular characteristics of their formation environment. Comparative planetology of the four giants in the solar system is therefore essential to reveal the potential formational, migrational, and evolutionary processes at work during the early evolution of the early solar nebula.

Much of our understanding of the origin and evolution of the outer planets comes from remote sensing by necessity. However, the efficiency of this technique has  limitations when used to study the bulk atmospheric composition that is crucial to the understanding of planetary origin, namely due to degeneracies between the effects of temperatures, clouds and abundances on the emergent spectra, but also due to the limited vertical resolution. In addition, many of the most common elements are locked away in a condensed phase in the upper troposphere, hiding the main volatile reservoir from the reaches of remote sensing.  It is only by penetrating below the ``visible'' weather layer that we can sample the deeper troposphere where those most common elements are well mixed. A remarkable example of the superiority of {\it in situ} probe measurements is illustrated by the exploration of Jupiter, where key measurements such as the determination of the noble gases abundances and the precise measurement of the helium mixing ratio have only been possible through {\it in situ} measurements by the Galileo probe \citep{1999Natur.402..269O}. 

The Galileo probe measurements provided new insights into the formation of the solar system. For instance, they revealed the unexpected enrichments of Ar, Kr and Xe with respect to their solar abundances, which suggested that the planet accreted icy planetesimals formed at temperatures possibly as low as 20--30 K to allow the trapping of these noble gases. Another remarkable result was the determination of the Jovian helium abundance using a dedicated instrument aboard the Galileo probe \citep{1998JGR...10322815V} with an accuracy of 2\%. Such an accuracy on the He/H$_2$ ratio is impossible to derive from remote sensing, irrespective of the giant planet being considered, and yet precise knowledge of this ratio is crucial for the modelling of giant planet interiors and thermal evolution. The Voyager mission has already shown that these ratios are far from being identical, which presumably results from slight differences in their histories at different heliocentric distances. An important result also obtained by the mass spectrometer onboard the Galileo probe was the determination of the $^{14}$N/$^{15}$N ratio, which suggested that nitrogen present in Jupiter today originated from the solar nebula essentially in the form of N$_2$ \citep{2001ApJ...553L..77O}. The Galileo science payload unfortunately could not probe to pressure levels deeper than 22 bars, precluding the determination of the H$_2$O abundance at levels representative of the bulk oxygen enrichment of the planet. Furthermore, the probe descended into a region depleted in volatiles and gases by unusual ``hot spot'' meteorology \citep{1998JGR...10322791O,2004Icar..171..153W}, and therefore its measurements are unlikely to represent the bulk planetary composition. Nevertheless, the Galileo probe measurements were a giant step forward in our understanding of Jupiter. However, with only a single example of a giant planet measurement, one must wonder whether from the measured pattern of elemental and isotopic enrichments, the chemical inventory and formation processes at work in our solar system are truly understood. {\it In situ} exploration of giant planets is the only way to firmly characterize the planet compositions in the solar system. In this context, a Saturn probe is the next natural step beyond Galileo's {\it in situ} exploration of Jupiter, the remote investigation of its interior and gravity field by the JUNO mission, and the Cassini spacecraft's orbital reconnaissance of Saturn. 

{\it In situ} exploration of Saturn's atmosphere addresses two broad themes. First, the formation history of our solar system and second, the processes at play in planetary atmospheres.  Both of these themes are discussed throughout this paper. Both themes have relevance far beyond the leap in understanding gained about an individual giant planet: the stochastic and positional variances produced within the solar nebula, the depth of the zonal winds, the propagation of atmospheric waves, the formation of clouds and hazes and disequilibrium processes of photochemistry and vertical mixing are common to all planetary atmospheres, from terrestrial planets to gas and ice giants and from brown dwarfs to hot exoplanets. 

This paper describes the main scientific goals to be addressed by the future {\it in situ} exploration of Saturn placing the Galileo probe exploration of Jupiter in a broader context and before the future {\it in situ} exploration of the more remote ice giants. These goals will become the primary objectives listed in the forthcoming Saturn probe proposals that we intent to submit in response to future opportunities within both ESA and NASA. Section \ref{origin} is devoted to a comparison between known elemental and isotopic compositions of Saturn and Jupiter. We describe the different formation scenarios that have been proposed to explain Jupiter's composition and discuss the key measurements at Saturn that would allow disentangling these interpretations. We also demonstrate that the {\it in situ} measurement of CO (or any other disequilibrium species that is depleted by reaction with water) at Saturn could place limits on its bulk O/H ratio. In Section \ref{atm}, we discuss the motivation for the {\it in situ} observation of the atmospheric processes (dynamics, chemistry and cloud formation) at work in Saturn's atmosphere. Section \ref{archi} is dedicated to a short description of the mission designs that can be envisaged. In Section \ref{inst}, we provide a description of high-level specifications for the science payload. Conclusions are given in Section \ref{cls}.

\section{Elemental and Isotopic Composition as a Window on Saturn's Formation}
\label{origin}

The giant planets in the solar system formed 4.55 Gyr ago from the same material that engendered the Sun and the entire solar system.  The envelopes of giant planets are dominated by hydrogen and helium, the two most abundant elements in the Universe. Protoplanetary disks, composed of gas and dust, are almost ubiquitous when stars form, but their typical lifetimes do not exceed a few million years. This implies that the gas giants Jupiter and Saturn had to form rapidly to capture their hydrogen and helium envelopes, more rapidly than the tens of millions of years needed for terrestrial planets  to reach their present masses. Due to formation at fairly large radial distances from the Sun, where the solid surface density is low, the ice giants Uranus and Neptune had longer formation timescales (slow growth rates) and did not manage to capture large amounts of hydrogen and helium before the disk gas dissipated. As a result, the masses of their gaseous envelopes are small compared to their
  ice/rock cores. 

A comparative study of the properties of these giant planets thus gives information on spatial gradients in the physical/chemical properties of the solar nebula as well as on stochastic effects that led to the formation of the solar system. Data on the composition and structure of the giant planets, which hold more than 95\% of the non-solar mass of the solar system, remain scarce, despite the importance of such knowledge. The formation of giant planets is now largely thought to have taken place via the core accretion model in which a dense core is first formed by accretion and the hydrogen-helium envelope is captured after a critical mass is reached \citep{1980PThPh..64..544M,1996Icar..124...62P}. When the possibility of planet migration is included \citep{1986ApJ...309..846L,1997Icar..126..261W}, such a model can explain the orbital properties of exoplanets, although lots of unresolved issues remain \citep{2004ApJ...604..388I,2012AA...547A.111M}. However, an alternative scenario for the formation of giant planets is the disk instability model \citep{1997Sci...276.1836B,2001ApJ...551L.167B}, in which the giant planets form from the direct contraction of a gas clump resulting from local gravitational instability in the disk. In principle, measurements of atmospheric bulk elemental enrichments and isotopic ratios would help us to distinguish between these competing formation scenarios. 

Formation and evolution models indicate that the total mass of heavy elements present in Jupiter may be as high as 42 \Mearth,~whereas the mass of the core is estimated to range between 0 and 13 \Mearth~\citep{2004ApJ...609.1170S}. In the case of Saturn, the mass of heavy elements can be as large as 35 \Mearth~with a mass varying between 0 and 10 \Mearth~in the envelope and the core mass ranging between 0 and 20 \Mearth~\citep{2013ApJ...767..113H}. Direct access to heavy materials within giant planet cores to constrain these models is impossible, so we must use the composition of the well-mixed troposphere to infer the properties of the deep interiors. It is difficult for remote sounding to provide the necessary information because of a lack of sensitivity to the atmospheric compositions beneath the cloudy, turbulent and chaotic weather layer. These questions must be addressed by {\it in situ} exploration, even if the NASA JUNO mission will try to address them remotely.

The availability of planetary building blocks (metals, oxides, silicates, ices) is expected to vary with position within the original nebula, from refractories in the warm inner nebula to a variety of ices of water, CH$_4$, CO, NH$_3$, N$_2$ and other simple molecules in the cold outer nebula. Turbulent radial mixing, and the evolution of the pressure-temperature gradient in the disk could have led to distinct regions where some species dominated over others (e.g., the water ice snowline or N$_2$ over NH$_3$). Furthermore, both inward and outward migration of the giants during their evolution could have provided access to different material reservoirs at different epochs. A giant planet's bulk composition therefore depends on the timing and location of planet formation, subsequent migration and the delivery mechanisms for the heavier elements. By measuring a giant planet's chemical inventory, and contrasting it with measurements of (i) other giant planets, (ii) primitive materials found in comets and asteroids, and (iii) the elemental abundances of our parent star and the local interstellar medium, we can reveal much about the conditions at work during the formation of our planetary system. 

It should be noted, however, that when atmospheric measurements are used to infer the planetary composition and reveal information on the planet's origin, one has to {\it assume} that the atmospheric composition represents the bulk planetary composition. This is a fairly good assumption if the measurements probe the convective region, and if the planet is fully convective. Within a fully convective planet the materials are expected to be homogeneously mixed, and therefore, we do not expect large differences in composition with depth. However, if the planet is not fully convective and homogeneously mixed, the information of its atmospheric composition cannot solely be used to infer the bulk composition.

In the case of Saturn (as well as Jupiter) compositional inhomogeneities can be the outcome of the formation process \citep[e.g.][]{1996Icar..124...62P} and/or the erosion of a primordial core that could mix with the surrounding metallic hydrogen \citep{2004PhT....57d..63G,2011APS..MARP31010W,2012PhRvL.108k1101W}. In addition, it is possible that double diffusive convection occurs in the interiors of giant planets \citep[e.g.][]{2012A&A...540A..20L,2013NatGe...6..347L}. If a molecular weight gradient is maintained throughout the planetary envelope, double-diffusive convection would take place, and the thermal structure would be very different from the one that is generally assumed using adiabatic (i.e., fully convective) models, with much higher center temperatures and a larger fraction of heavy elements. In this case, the planetary composition can vary substantially with depth and therefore, a measured composition of the envelope would not represent the overall composition. While standard interior models of Saturn assumed three layers and similar constraints in terms of the helium to hydrogen ratio, they can differ in the assumption on the distribution of heavy elements within the planetary envelope. While Guillot and collaborators \citep[e.g.][]{2004ApJ...609.1170S,2013ApJ...767..113H} assume homogeneous distribution of heavy elements apart from helium, which is depleted in the outer envelope due to helium rain (a process that seems to be related also to neon depletion), interior structure models by Nettelmann and collaborators \citep{2010SSRv..152..423F,2013Icar..225..548N} allow the abundance of heavy elements to be discontinuous between the molecular and the metallic envelope. At present, it is not clear whether there should be a discontinuity in the composition of heavy elements, and this question remains open.

\subsection{Jupiter and Saturn's Composition}
\label{comp}

The abundances and isotopic ratios of most significant volatiles measured at Jupiter and Saturn are given in Tables 1 and 2. We refer the reader to the papers of \cite{2003PSS...51..105A}, \cite{2006Icar..185..466T} and \cite{2012A&A...539A..44F} for a more exhaustive list of disequilibrium species identified (or for other minor species presumably identified) in Jupiter's and Saturn's atmospheres. Only upper limits on the abundances of hydrogen halides have been derived from the remote detection of these species in Saturn's atmosphere, implying the need of a probe to get improved {\it in situ} measurements.

The abundances of CH$_4$, NH$_3$, H$_2$O, H$_2$S, Ne, Ar, Kr and Xe have been measured by the Galileo Probe Mass Spectrometer (GPMS) in Jupiter's atmosphere \citep{2000JGR...10515061M,2004Icar..171..153W}. The value of H$_2$O abundance reported for Jupiter in Table 1 corresponds to the deepest measurement made by the probe (at 17.6--20.9 bar) and is probably much smaller than the planet's bulk water abundance, which remains unknown \citep{2003PSS...51..105A,2004Icar..171..153W}. The Juno mission, which will arrive at Jupiter in 2016, may provide an estimate of the tropospheric O/H ratio. The He abundance in Jupiter has also been measured {\it in situ} by a Jamin-Mascart interferometer aboard the Galileo probe (Helium Abundance Detector; hereafter HAD) with a better accuracy level than the GPMS instrument \citep{1998JGR...10322815V}. PH$_3$ is the only species of our list of Jupiter measurements whose abundance has been determined remotely by the Cassini Composite Infrared Spectrometer (CIRS) during the spacecraft 2000--2001 encounter \citep{2009Icar..202..543F}. PH$_3$ is a disequilibrium species at its sampling level in Jupiter's atmosphere (see Sec.\ 3), implying that its bulk abundance is probably higher than the measured one. Isotopic measurements presented for Jupiter in Table 2 have also been performed by the GPMS instrument aboard the Galileo probe \citep{1996Sci...272..846N,1998JGR...10322831N,2000JGR...10515061M,2003PSS...51..105A,2004Icar..171..153W}.

In the case of Saturn, only the abundances of CH$_4$, PH$_3$, NH$_3$ and H$_2$O, and indirectly that of H$_2$S, have been measured. The abundance of CH$_4$ has been determined from the analysis of high spectral resolution observations from CIRS \citep{2009Icar..199..351F}. Similarly to Jupiter, PH$_3$  has been determined remotely in Saturn from Cassini/CIRS observations at 10 $\mu$m \citep{2009Icar..202..543F}. Other measurements of PH$_3$ have been made from ground based observations at 5 $\mu$m \citep{1997AA...321L..13D}, but the spectral line data at these wavelengths is less robust and accurate than those at 10 $\mu$m. There is also a degeneracy with the location, extent, opacity of Saturn's clouds at 5 $\mu$m which is not apparent at 10 $\mu$m. Moreover, considering the fact that there is also terrestrial contamination in the 5 $\mu$m window for groundbased observations and that the scattered sunlight may contribute at 5 $\mu$m, this leads us to believe that the data at 10 $\mu$m are more reliable. Interestingly, we note that PH$_3$ is easier to detect on Saturn compared to Jupiter because this molecule dominates the upper tropospheric chemistry and ammonia is locked away at deeper levels. The NH$_3$ abundance is taken from the range of values derived by \cite{2011Icar..214..510F} who analyzed Saturn's tropospheric composition from Cassini/VIMS 4.6--5.1 $\mu$m thermal emission spectroscopy. This determination is probably more reliable than those made in the microwave domain because of the absence of spectral lines at these wavelengths \citep{1989Icar...80...77B,2013Icar..226..641L}. Tropospheric H$_2$O has been inferred in Saturn via the Short Wavelength Spectrometer Instrument onboard the Infrared Space Observatory (ISO-SWS) \citep{1997AA...321L..13D}. However, H$_2$O is unsaturated at this altitude ($\sim$3 bar level), implying that its bulk abundance is higher than the measured one. The H$_2$S abundance is quoted from the indirect determination of \cite{1989Icar...80...77B} who investigated the influence of models of NH$_3$-H$_2$S-H$_2$O cloud decks on Saturn's atmospheric opacity at microwave wavelengths. The He abundance in Saturn's atmosphere derives from a reanalysis of Voyager's infrared spectrometer (IRIS) measurements \citep{2000Icar..144..124C}.  The only isotopic ratios measured in Saturn are D/H in H$_2$ \citep[determination from ISO-SWS,][]{2001AA...370..610L} and $^{12}$C/$^{13}$C in CH$_4$ \citep[Cassini/CIRS observations,][]{2009Icar..199..351F}.

Table 3 summarizes the enrichments in volatiles relative to protosolar values observed in Jupiter and Saturn.  Note that protosolar abundances are different from present-day solar photospheric abundances because elements heavier than He are settling out of the photosphere over time. This mechanism leads to a fractionation of heavy elements relative to hydrogen in the solar photosphere, requiring the use of correction terms to retrieve the protosolar abundances \citep{2009ARA&A..47..481A}. For the sake of information, the protosolar elemental abundances used in our calculations are detailed in Table 4. C, N, P, S, Ar, Kr and Xe are all found enriched by a factor $\sim$2 to 4 in Jupiter. On the other hand, C, N and P (the only heavy elements {\it a priori} reliably measured) are found enriched by factors of $\sim$10, 0.5--5 and 11.5 in Saturn. Helium is depleted compared to protosolar values in the two giants because of its condensation into droplets that ``rain out'' in the giant planets deep interiors \citep{1977ApJS...35..239S,1977ApJS...35..221S,2003Icar..164..228F}. The solution of neon in those droplets \citep{2010PhRvL.104l1101W} would also explain its apparent depletion in Jupiter but a similar measurement has never been possible on Saturn. As mentioned above, oxygen is also depleted compared to protosolar in the Jovian atmosphere but this measurement results from the fact that the Galileo probe entry site was an unusually dry meteorological system. As a result, the probe did not measure the deep, well-mixed water mixing ratio \citep{2004Icar..171..153W}, which is predicted to be supersolar \citep{1988Icar...75..146S,2001ApJ...550L.227G,2004P&SS...52..623H,2005ApJ...622L.145A,2009ApJ...696.1348M,2012ApJ...751L...7M}.

\subsection{Indirect Determination of Saturn's O/H Ratio}
\label{OtoH}

One of the main objectives of Saturn's {\it in situ} exploration is the measurement of the H$_2$O abundance. However, depending on the O/H elemental enrichment \citep{1999P&SS...47.1243A}, H$_2$O is predicted to condense in the 12.6--21 bar range and may remain out of reach for the probe we consider in this paper that would be limited to $\sim$10 bar (see Sec. \ref{archi}). Several disequilibrium species, like CO, can provide useful constraints on Saturn's deep H$_2$O abundance. The upper tropospheric mole fraction of CO is representative of the H$_2$O abundance in the deep hot troposphere, where the two species are in thermochemical equilibrium \citep{1994Icar..110..117F}. It is thus possible to derive the deep H$_2$O abundance from CO observations using the ``quench level''  approximation (e.g., \citealt{2002Icar..159...95B}), or more rigorously using comprehensive thermochemical models (e.g., \citealt{2010Icar..209..602V} and \citealt{Cavalie2014}).

\begin{figure}[h]
\begin{center}
\resizebox{\hsize}{!}{\includegraphics[angle=0]{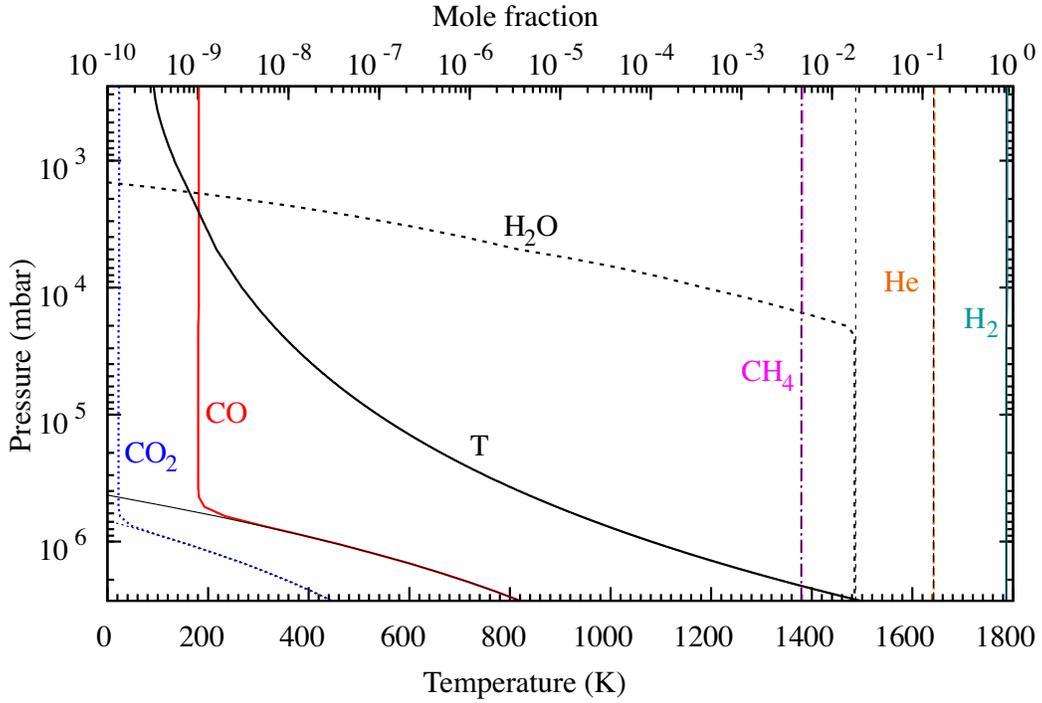}}
\end{center}
\caption{Mole fraction profiles in the troposphere of Saturn obtained with \cite{2012A&A...546A..43V}'s model, targeting the $10^{-9}$ upper limit on the upper tropospheric CO mole fraction obtained by \cite{2009Icar..203..531C}. The temperature profile in the troposphere is shown in black solid line. Thermochemical equilibrium profiles are shown as black solid lines with the same layout as their corresponding species. The model parameters are: O/H$=$ 21 times solar, C/H$=$ 9 times solar, and $K_{zz}$$~=~$$ 10^9$\,\Kunit. Condensation of H$_2$O occurs around the 20 bar level in this model.}
\label{Saturn_Kzz1d9} 
\end{figure}

We have adapted the model of \cite{2012A&A...546A..43V} to Saturn's troposphere to assess the relevance of measuring CO with an \textit{in situ} probe. The thermochemical kinetic network comes from the engine industry and was thoroughly validated for high temperatures and pressures. The tropospheric thermal profile has been constructed from a recent retrieval of the latitudinally-resolved $T$($P$) structure representing a mean of Cassini's prime mission \citep{2009Icar..199..351F}. We used the nominal mixing ratios from Table 1 for He and CH$_4$, and adopted an upper limit of $10^{-9}$ for CO \citep{2009Icar..203..531C}. We have assumed a vertically constant eddy mixing coefficient $K_{zz}$ ranging from $10^{8}$ to $10^{9}$\,\Kunit~\citep{2010Icar..209..602V}. With $K_{zz}$$=$$10^{8}$\,\Kunit, the deep atmospheric O/H ratio needs to be 62 times the protosolar value to reproduce the CO upper limit. With $K_{zz}$$=$$10^{9}$\,\Kunit, the O/H still needs to be 18 times protosolar (see \fig{Saturn_Kzz1d9}), i.e., still much higher than Saturn's C/H ratio (9.9 times protosolar) but remains within the range of values predicted from the theory arguing that volatiles formed clathrates and pure condensates in the nebula (see Sec.\ 2.3.2). If we reversely set O/H ratio to the C/H one, then the most favorable case for a detection of CO ($K_{zz}$$=$$10^{9}$\,\Kunit) gives an upper tropospheric mole fraction of CO of 4.1\dix{-10}. Reaching such a low value will remain very challenging for any ground-based facility. Besides, a complication comes from the fact that the observable CO vertical profile is largely dominated by an external source in the stratosphere \citep{2010A&A...510A..88C}.

These results argue in favor of an \textit{in situ} measurement of tropospheric CO with a neutral mass spectrometer as a valuable complement to any attempt to directly measure the H$_2$O abundance. However, CO has a molecular weight very close to that of N$_2$. This degeneracy is a serious issue because the N$_2$ upper tropospheric mole fraction is expected to be around four orders of magnitude higher than the one of CO. A mass spectrometer will therefore need a mass resolution of $m/\Delta m = 2,500$ to separate CO from N$_2$ at equal abundance, and about $m/\Delta m = 15,000$ for the CO and N$_2$ abundances expected in Saturn's atmosphere. More generally, any other disequilibrium species that reacts with H$_2$O, like PH$_3$ and SiH$_4$, is likely to provide additional constraints on the deep H$_2$O abundance of Saturn \citep{2005ApJ...623.1221V} and it would be desirable to include the combustion reaction schemes of such species (e.g., \citealt{Twarowski1995} and \citealt{Miller2004}) in thermochemical models.

\subsection{Isotopic Measurements at Saturn}
\label{Isotop}

As shown in Table 2, very little is known today concerning the isotopic ratios in Saturn's atmosphere. Only D/H (for H$_2$ and methane) and $^{12}$C/$^{13}$C (for methane) ratios have been measured so far \citep{2001AA...370..610L,2009Icar..199..351F}.

The case of D/H is interesting and would deserve further measurements with smaller errors. Because deuterium is destroyed in stellar interiors and transformed into $^3$He, the D/H value presently measured in Jupiter's atmosphere is estimated to be larger by some 5--10\% than the protosolar value. This slight enrichment would have resulted from a mixing of nebular gas with deuterium-rich ices during the planet's formation, as suggested by \cite{1999P&SS...47.1183G}. For Saturn, the contribution of deuterium-rich ices in the present D/H ratio could be higher (25--40\%). An accurate measurement of the D/H ratio in Saturn's atmosphere could provide, consequently, some constraints on the relative contribution of deuterium-rich ices during the formation of Saturn. Such a constraint is also based on the {\it a priori} knowledge of the protosolar D/H ratio, which remains relatively uncertain. This ratio is estimated from measurements of $^3$He/$^4$He in the solar wind, which is corrected for changes that occurred in the solar corona and chromosphere subsequently to the evolution of the Sun's interior, and to which the primordial $^3$He/$^4$He is subtracted. This latter value is estimated from the ratio observed in meteorites or in Jupiter's atmosphere. The measurement of $^3$He/$^4$He in Saturn's atmosphere would also complement, consequently, the scientific impact of D/H measurement. In any case the smaller value of D/H measured by \cite{2001AA...370..610L} in Saturn's atmosphere from infrared spectra obtained by the Infrared Space Observatory (ISO) satellite and the Short Wavelength Spectrometer (SWS) compared to Jupiter's atmosphere \citep{1998JGR...10322831N} is surprising in the sense that it would suggest a lower relative contribution of deuterium-rich ices in the formation of Saturn compared to Jupiter. These values have, nevertheless, large errors and so far no clear conclusion can be drawn.

The $^{14}$N/$^{15}$N ratio presents large variations in the different planetary bodies in which it has been measured and, consequently, remains difficult to interpret. The analysis of Genesis solar wind samples \citep{2011Sci...332.1533M} suggests a $^{14}$N/$^{15}$N ratio of $441\pm5$, which agrees with the {\it in situ} measurements made in Jupiter's atmospheric ammonia \citep{2000Icar..143..223F,2004Icar..172...50F} which probably comes from primordial N$_2$ \citep{2001ApJ...553L..77O}. Terrestrial atmospheric N$_2$, with a value of 272, appears enriched in $^{15}$N compared to Jupiter and similar to the bulk of ratios derived from the analysis of comet 81P/ Wild 2 grains \citep{2006Sci...314.1724M}. Measurements performed in Titan's atmosphere, which is dominated by N$_2$  molecules, lead to $167.7\pm0. 6$ and $147.5\pm7.5$  from the Cassini/INMS and Huygens/GCMS data, respectively \citep{2010JGRE..11512006N,2009P&SS...57.1917M}. Because of the low abundance of primordial Ar observed by Huygens, it is generally assumed that N$_2$ is of secondary origin in Titan's atmosphere and that N was delivered in a less volatile form, probably NH$_3$. Different mechanisms have been proposed for the conversion of NH$_3$  to N$_2$. Isotopic fractionation may have occurred for nitrogen in Titan's atmosphere but the atmospheric model published by \cite{2009P&SS...57.1917M} suggests that the current $^{14}$N/$^{15}$N ratio observed in N$_2$  is close to the value acquired by the primordial ammonia of Titan. This statement is supported by the recent measurement of the $^{14}$N/$^{15}$N isotopic ratio in cometary ammonia \citep{2014ApJ...780L..17R}. This ratio, comprised between 80 and 190, is consistent with the one measured in Titan.

All these measurements suggest that N$_2$  and NH$_3$ result from the separation of nitrogen into at least two distinct reservoirs, with a distinct $^{15}$N enrichment, which never equilibrated. The reservoir containing N$_2$ would have a large $^{14}$N/$^{15}$N ratio (like in Jupiter's atmosphere, where the present ammonia is supposed to come from primordial N$_2$) and the one containing NH$_3$ a much lower value (like in Titan's atmosphere, where the present N$_2$ could come from primordial ammonia, and in cometary ammonia). In this context measuring $^{14}$N/$^{15}$N in Saturn's atmosphere would be very helpful to get more information about the origin of ammonia in this planet.

The cases of carbon, oxygen and noble gas (Ne, Ar, Kr, and Xe) isotopic ratios are different because they should be representative of their primordial values. Only little variations are observed for the $^{12}$C/$^{13}$C ratio in the solar system irrespective of the body and molecule in which it has been measured. This ratio appears compatible with the terrestrial value of 89 (except if isotopic fractionation processes occur, like for methane in Titan, but the influence of these processes on this ratio is small). Table 2 provides the value of 91.8 measured by \cite{2009Icar..199..351F} in Saturn with the Cassini/CIRS but with large error bars. A new {\it in situ} measurement of this ratio should be useful to confirm that carbon in Saturn is also representative of the protosolar value (and different from the one present in the local Interstellar Medium (ISM) because $^{13}$C is created in stars). The oxygen isotopic ratios also constitute interesting measurements to be made in Saturn's atmosphere. The terrestrial $^{16}$O/$^{18}$O and $^{16}$O/$^{17}$O isotopic ratios are 499 and 2632, respectively \citep{2009ARA&A..47..481A}. At the high accuracy levels possible with meteorites analysis these ratios present some small variations\footnote{Expressed in $\delta$ units, which are deviations in part per thousand, they are typically a few units.}. Measurements performed for solar system objects like comets, far less accurate, match the terrestrial $^{16}$O/$^{18}$O value (with error bars being typically a few tens). However no $^{16}$O/$^{18}$O ratio has been yet published for Saturn's atmosphere. The only $^{16}$O/$^{18}$O measurement made so far for a giant planet \citep{1995ApJ...453L..49N} was obtained from groundbased IR observations in Jupiter's atmosphere and had a very large uncertainty (1--3 times the terrestrial value). 

\subsection{Interpretations of the Volatile Enrichments in Jupiter and Saturn}
\label{interp}

Several theories connecting the thermodynamic evolution of the protosolar nebula to the formation conditions of the giant planets have been developed to interpret the volatile enrichments measured in Jupiter and Saturn. The main scenarios proposed in the literature and their predictions for Saturn's composition are summarized below.

\subsubsection{Amorphous Ice Scenario}
\label{sc1}

The model proposed by \cite{1999Natur.402..269O} is the first attempt to explain of the volatile enrichments measured in Jupiter's atmosphere. In this scenario, volatiles present in Jupiter's atmosphere were first acquired in amorphous ice at temperatures as low as 30 K in the protosolar nebula. This hypothesis is based on the fact that formation scenarios of the protosolar nebula invoke two reservoirs of ices, namely an inner and an outer reservoir, that took part in the production of icy planetesimals. The first reservoir, located within $\sim$30 Astronomical Units (AU) of the Sun, contains ices (mostly water ice) originating from the ISM which, due to their proximity to the Sun, were initially vaporized \citep{1997ApJ...477..398C}. With time, the decrease of temperature and pressure conditions allowed the water in this reservoir to condense at $\sim$150 K at nebular pressure conditions in the form of (microscopic) crystalline ice \citep{1994A&A...290.1009K}. The other reservoir, located at larger heliocentric distances, is composed of ices originating from the ISM that did not vaporize when entering into the disk. In this reservoir, water ice was essentially in the amorphous form and the other volatiles remained trapped in the amorphous matrix \citep{2005Icar..175..546N}. In this context, to explain the accretion of amorphous planetesimals by the forming Jupiter, \cite{1999Natur.402..269O} proposed that either the giant planet was formed at large heliocentric distances where the temperature always favored the preservation of amorphous ice in the disk, or the protosolar nebula was much cooler at the current location of Jupiter ($\sim$5 AU) than predicted by current turbulent accretion disk models. In both cases, the icy material originated from the protosolar cloud and survived the formation of the protosolar nebula. This is the fraction of the icy planetesimals that vaporized when entering the envelopes of the growing Jupiter, which engendered the observed volatile enrichments. If correct, this scenario predicts that the volatile enrichments at Saturn should be in solar proportions, as seems to be the case in Jupiter, given the size of the error bars of measurements.

\subsubsection{Crystalline Ice Scenario}
\label{sc2}

An alternative interpretation of the volatile enrichments measured in Jupiter is the one proposed by \cite{2001ApJ...550L.227G} and subsequent papers by \cite{2004P&SS...52..623H}, \cite{2005ApJ...622L.145A} and \cite{2006A&A...449..411M}. These authors assumed that Jupiter's building blocks formed in the inner 30 AU of the disk, in which the gas phase has been enriched at early epochs by the vaporization of amorphous ice entering from the ISM. During the cooling of this region of the disk, water vapor crystallized and trapped the volatiles in the form of clathrates or hydrates in the 40--90 K range instead of condensing at lower temperatures. These ices then agglomerated and formed the planetesimals that were ultimately accreted by the growing Jupiter. These scenarios postulate that the amount of available crystalline water ice was large enough (typically H$_2$O/H$_2$~$\ge$~2~$\times$~(O/H)$_\odot$) to trap the other volatiles in the feeding zone of Jupiter and that the disk's temperature at which the ices formed never decreased below $\sim$40 K. 

Subsequent works have shown that it is possible to explain the volatile enrichments in Jupiter via the accretion and the vaporization in its envelope of icy planetesimals made from a mixture of clathrates and pure condensates \citep{2009ApJ...696.1348M,2012ApJ...751L...7M}. These planetesimals could have formed if the initial disk's gas phase composition was fully protosolar (including oxygen), and if the disk's temperature decreased down to $\sim$20 K at their formation location. The model described in \cite{2009ApJ...696.1348M,2012ApJ...751L...7M} is used here to show fits of the volatile enrichments measured at Jupiter and Saturn, which have been updated by using the recent protosolar abundances of \cite{2009LanB...4B...44L} (see Table 3). With this model, we first computed the composition of planetesimals condensed from two extreme gas phase compositions of the nebula, namely oxidizing and reducing states. In the oxidizing state, oxygen, carbon, nitrogen are postulated to exist only in the molecular species H$_2$O, CO, CO$_2$, CH$_3$OH, CH$_4$, N$_2$, and NH$_3$. We fixed CO/CO$_2$/CH$_3$OH/CH$_4$ = 70/10/2/1 and N$_2$/NH$_3$ = 10 in the gas phase of the disk, values usually used for the protosolar nebula \citep{2009ApJ...696.1348M,2012ApJ...751L...7M}. In contrast, in the reducing state,  C exists only in CH$_4$ form and N$_2$/NH$_3$ = 0.1 in the gas phase \citep{2012ApJ...757..192J}. In both cases, P is in the form of PH$_3$ and the volatile fraction of S is assumed to exist in the form of H$_2$S, with H$_2$S/H$_2$~=~0.5~$\times$~(S/H$_2$)$_{\odot}$, the other fraction of S being essentially trapped in the form of troilite mineral in the solar nebula \citep{2005Icar..175....1P}. Planetesimals formed during the cooling of the nebula from these two extreme gas phase compositions are assumed to have been accreted by proto-Jupiter and proto-Saturn and devolatilized in the envelopes during their growth phases. Once the composition of the planetesimals is defined, the adjustment of their masses accreted in the envelopes of Jupiter and Saturn allows one to determine the best fit of the observed volatile enrichments.

\begin{figure}[!h]
\begin{center}
\resizebox{\hsize}{!}{\includegraphics[angle=0]{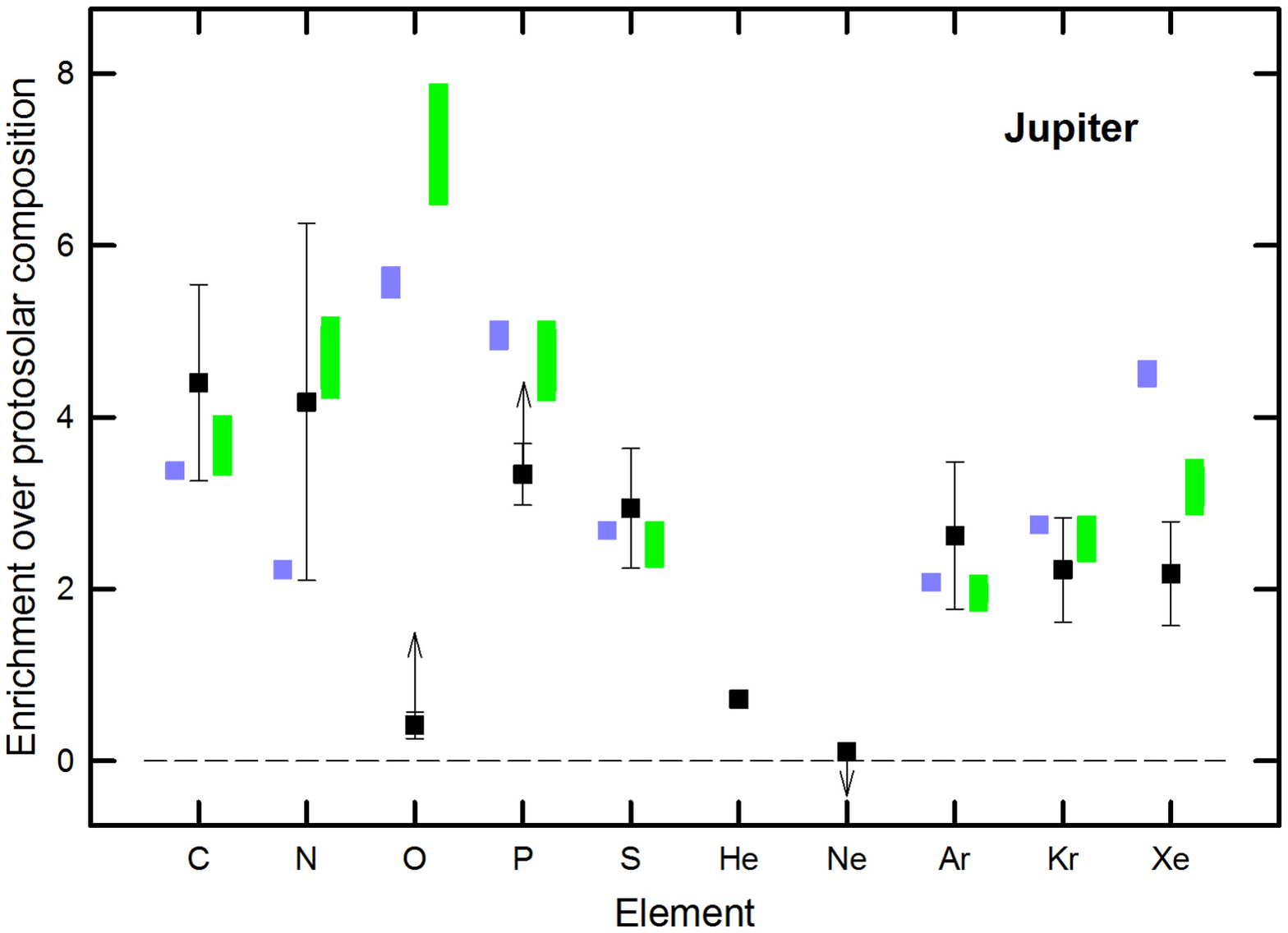}}
\end{center}
\caption{Ratio of Jovian to protosolar abundances. Black squares and black bars correspond to measurements and their associated uncertainties. Blue and green bars correspond to calculations assuming oxidizing and reducing conditions in the protosolar nebula, respectively (see text). Arrows pointing up correspond to the possibility that the measured oxygen and phosphorus abundances are lower than their bulk abundances, and arrow pointing down to the fact that the measured Ne abundance is an upper limit.}
\label{Jupiter_fits} 
\end{figure}

\begin{figure}[!h]
\begin{center}
\resizebox{\hsize}{!}{\includegraphics[angle=0]{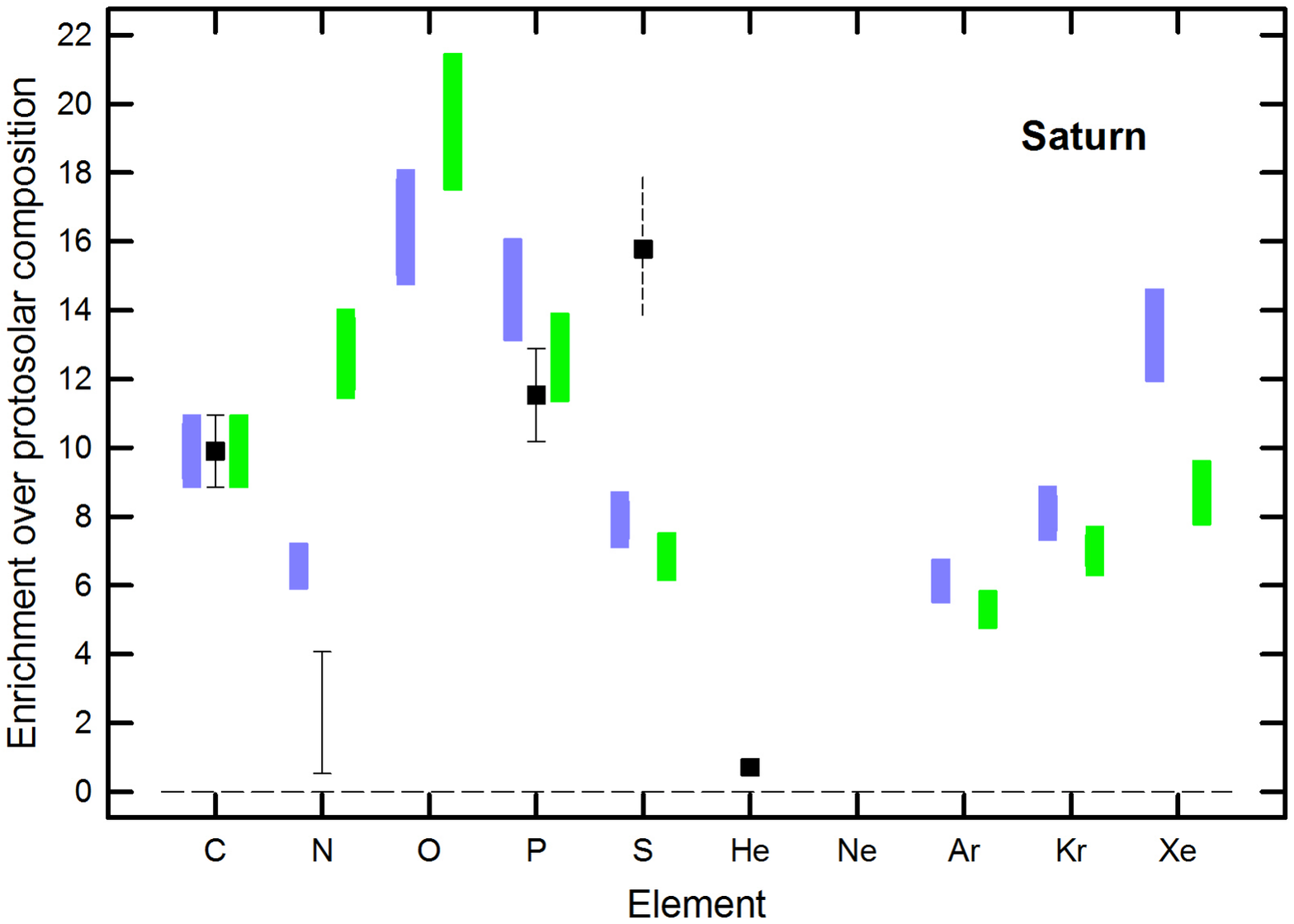}}
\end{center}
\caption{Ratio of Saturnian to protosolar abundances. Black squares and black bars correspond to measurements and their associated uncertainties. The O value measured in the troposphere would be close to zero on the utilized scale. Blue and green bars correspond to calculations assuming oxidizing and reducing conditions in the protosolar nebula, respectively (see text).}
\label{Saturn_fits} 
\end{figure}

Figures \ref{Jupiter_fits} and \ref{Saturn_fits} represent the fits of the enrichments observed in Jupiter's and Saturn's atmospheres, respectively. In the case of Jupiter, C, N, S, Ar and Kr measurements are matched by our fits, irrespective of the redox status of the protosolar nebula. Also, in both redox cases, the measured P abundance is not matched by the fits but this might be due to the difficulty of getting a reliable measurement since the mid-infrared spectrum is dominated by tropospheric ammonia. Also the measured P is predicted to be lower than its bulk abundance due to disequilibrium processes in the Jovian atmosphere \citep{2009Icar..202..543F}. On the other hand, Xe is almost matched by our fit in the reducing case only. The oxygen abundance is predicted to be 5.4--5.7 and 6.5--7.9 times protosolar in Jupiter in the oxidizing and reducing cases, respectively.

In the case of Saturn, our strategy was to fit the measured C enrichment. Interestingly, contrary to Jupiter, P is matched in Saturn, irrespective of the redox status of the nebula. On the other hand, the P determination is more robust in Saturn than in Jupiter because PH$_3$ dominates the mid-infrared spectrum. However, S is not matched by our model but this might result from the lack of reliability of its determination. In addition, with enrichments predicted to be $\sim$6--7 times and 11--14 times the protosolar value in the oxidizing and reducing cases, respectively, our model overestimates the amount of nitrogen present in Saturn's atmosphere compared to observations that suggest a more moderate enrichment, in the order of $\sim$0.5--4.6 times the protosolar value. One possibility that could explain this discrepancy is that all NH$_3$ and only a fraction of N$_2$ would have been incorporated in Saturn's building blocks because of the limited amount of available water favoring its efficient trapping in clathrates. The remaining fraction of N$_2$ would have remained in the H$_2$-dominated gas phase of Saturn's feeding zone as a result of the disk's cooling down to temperatures higher than that of N$_2$ condensation or trapping in clathrates, as proposed by \cite{2008P&SS...56.1103H}. These conditions could lead to a moderate N enrichment comparable to the measured one and to a $^{14}$N/$^{15}$N ratio in the envelope very close to the Jovian value. Our model also gives predictions of O, Ar, Kr and Xe enrichments in the two redox cases. In particular, the oxygen abundance is predicted to be 14.7--18.1 and 17.5--21.5 times protosolar in the oxidizing and reducing cases, respectively. If the determination of N is confirmed at Saturn, it would appear inconsistent with the scenario proposed by \cite{1999Natur.402..269O} because the latter predicts a uniform enrichment in volatiles in the giant planet's envelope, which is not the case here since C/N is found to be at least $\sim$2 $\times$ (C/N)$_\odot$. On the other hand, both scenarios predict the same $^{14}$N/$^{15}$N ratio at Jupiter and Saturn as the two planets accreted their nitrogen essentially from the same volatile reservoirs.

\subsubsection{Scenario of Supply of Refractory Carbonated Material }
\label{sc3}

\cite{2004ApJ...611..587L} proposed the formation of Jupiter from refractory carbonated materials, namely ``tar'', placing its formation location on a ``tar line'' in the protosolar nebula. This scenario was used to explain
the elemental abundances enrichments observed by Galileo after having normalized all the heavy elements abundances with respect to Si instead of H$_2$. By doing so, \cite{2004ApJ...611..587L} found that the relative abundances of Ar, Kr, Xe and P are solar, C and possibly N are enriched, and H, He, Ne, and O are subsolar, with the Galileo H$_2$O determination assumed to be representative of the planet's bulk O/H. In this model, Ar, Kr and Xe would have been supplied to Jupiter via direct gravitational capture of the solar nebula gas. To explain the Ar, Kr and Xe enrichments in the Jovian atmosphere, \cite{2004ApJ...611..587L} proposed that they would have been the consequence of the H$_2$ and He depletion in the envelope, which produced the metallic layer. If Saturn formed following this scenario, a useful test would be the determination of the H$_2$O bulk abundance, which should be subsolar, as proposed by \cite{2004ApJ...611..587L} for Jupiter.

\subsubsection{Scenario of Disk's Gas Phase Enrichment}
\label{sc4}

To account for the enrichments in heavy noble gases observed in Jupiter's atmosphere, \cite{2006MNRAS.367L..47G} proposed that Ar, Kr and Xe have condensed at $\sim$20--30 K onto the icy amorphous grains that settled in the cold outer part of the disk nebula midplane. These noble gases would have been released in gaseous form in the formation region of giant planets at a time when the disk would have been chemically evolved due to photoevaporation. The combination of these mechanisms would have led to a heavy noble gas enrichment relative to protosolar in the disk's gas phase from which the giant planets would have been accreted. In \cite{2006MNRAS.367L..47G}'s scenario, the noble gas enrichment would have been homogeneous in the giant planets formation region. Therefore, their model predicts that the Ar, Kr and Xe enrichments in Saturn's atmosphere are similar to those observed in Jupiter, which are between $\sim$1.8 and 3.5 times the protosolar value (see Table 3). These values are substantially smaller than those predicted by the model used in Sec. \ref{sc2}, which are in the $\sim$4.8--14.6 times protosolar range, depending on the considered species (see Fig. \ref{Saturn_fits}).

\subsection{Summary of Key Measurements}
\label{key}

Here we provide the ``recommended'' measurements in Saturn's atmosphere that would allow disentangling between i) the afore-mentioned giant planets formation scenarios and ii) the different volatile reservoirs from which the solar system bodies assembled:

\begin{itemize}
\item The atmospheric fraction of He/H$_2$ with a 2\% accuracy on the measurement (same accuracy as the one made by the Jamin-Mascart interferometer aboard Galileo);
\item The elemental enrichments in cosmogenically abundant species C, N, S and O. C/H, N/H, S/H and O/H should be sampled with an accuracy better than $\pm$ 10\% (uncertainties of the order of protosolar abundances).
\item The elemental enrichments in minor species delivered by vertical mixing (e.g., P, As, Ge) from the deeper troposphere (see also Sec. \ref{atm}). P/H, As/H and Ge/H should be sampled with an accuracy better than $\pm$ 10\% (uncertainties of the order of protosolar abundances).
\item The isotopic ratios in hydrogen (D/H), oxygen ($^{18}$O, $^{17}$O and $^{16}$O), carbon ($^{13}$C/$^{12}$C) and nitrogen ($^{15}$N/$^{14}$N), to determine the key reservoirs for these species (e.g., delivery as N$_2$ or NH$_3$ vastly alters the $^{15}$N/$^{14}$N ratio in the giant planet's envelope). $^{13}$C/$^{12}$C, $^{18}$O/$^{16}$O and $^{17}$O/$^{16}$O should be sampled with an accuracy better than $\pm$ 1\%. D/H, $^{15}$N/$^{14}$N should be analyzed in the main host molecules with an accuracy of the order of $\pm$ 5\%.
\item The abundances and isotopic ratios for the chemically inert noble gases He, Ne, Xe, Kr and Ar, provide excellent tracers for the materials in the subreservoirs existing in the protosolar nebula. The isotopic ratios for He, Ne, Xe, Kr and Ar should be measured with an accuracy better than $\pm$ 1\%.
\end{itemize}

The depth of probe penetration will determine whether it can access the well-mixed regions for key condensable volatiles. In the case of Saturn, a shallow probe penetrating down to 5--10 bar would {\it in principle} sample ammonia and H$_2$S both within and below their cloud bases, in the well-mixed regions of the atmosphere to determine the N/H and S/H ratios, in addition to noble gases and isotopic ratios. Note that the N determination could be a lower limit because ammonia is highly soluble in liquid water. Rain generated in the water cloud can provide a downward transport mechanism for ammonia, so the ammonia abundance above the water cloud could be less than the bulk abundance. Because the hypothesized water cloud is deeper than at least $\sim$12.6 bar in Saturn \citep{1999P&SS...47.1243A}, the prospect of reaching the deep O/H ratio remains unlikely if the probe would not survive beyond its design limit, unless a precise determination of the CO abundance (or any other species limited by reactions with the tropospheric water) is used to constrain H$_2$O/H$_2$ (see Sec. \ref{OtoH}) and/or the probe is accompanied by remote sensing experiments on a carrier spacecraft capable of probing these depths (e.g., the Juno microwave radiometer, currently en route to Jupiter). Nevertheless, measuring elemental abundances (in particular He, noble gases and other cosmogenically-common species) and isotopic ratios using a shallow entry probe on Saturn will provide a vital comparison to Galileo's measurements of Jupiter, and a crucial ``ground-truth'' for the remote sensing investigations by the Cassini spacecraft.

\section{\textit{In situ} Studies of Saturn's Atmospheric Phenomena}
\label{atm}

The giant planets are natural planetary-scale laboratories for the study of fluid dynamics without the complicating influences of terrestrial topography or ocean-atmosphere coupling.  However, remote sensing only provides access to limited altitude ranges where spectral lines are formed and broadened.  Furthermore, the vertical resolution of ``nadir'' remote sensing is fundamentally limited to the width of the contribution function (i.e., the range of altitudes contributing to the upwelling radiance at a given wavelength), which can extend over a broad range of pressures.  Ground-based observatories, space telescopes and the visiting Pioneer, Voyager and Cassini missions have exploited wavelengths from the ultraviolet to the microwave in an attempt to reconstruct Saturn's atmospheric structure in three dimensions. These studies have a limited vertical resolution and principally use visible and infrared observations in the upper troposphere (just above the condensate clouds and within the tropospheric hazes) or the mid-stratosphere near the 1 mbar level via mid-infrared emissions.  Regions below the top-most clouds and in the middle/upper atmosphere are largely inaccessible to remote sensing, limiting our knowledge of the vertical variations of temperatures, densities, horizontal and vertical winds and waves, compositional profiles and cloud/haze properties. \textit{In situ} exploration of Saturn would not only help constrain the bulk chemical composition of this gas giant (e.g., Section \ref{origin}), but it would also provide direct sampling and ``ground-truth'' for the myriad physical and chemical processes at work in Saturn's atmosphere.

In the following sections we describe how an \textit{in situ} probe, penetrating from the upper atmosphere ($\mu$bar pressures) into the convective weather layer to a minimum depth of 10 bar, would contribute to our knowledge of Saturn's atmospheric structure, dynamics, composition, chemistry and cloud-forming processes.  These results would be directly compared to our only other direct measurement of a giant planet, from the descent of the 339-kg Galileo probe into the atmosphere of Jupiter on December 7th 1995.  The Galileo probe entered a region of unusual atmospheric dynamics near 6.5$^\circ$N, where it is thought that the meteorology associated with planetary wave activity conspired to deplete Jupiter's atmosphere in volatiles \citep[e.g.,][]{2000Sci...289.1737S,1999Icar..137...34F}, most notably preventing the probe from reaching the depth of Jupiter's well-mixed H$_2$O layer after its 60-minute descent to the 22 bar level, 150~km below the visible cloud-tops.  In the decade that followed, researchers have been attempting to reconcile global remote sensing of Jupiter with this single-point measurement \citep[e.g.,][]{2000Natur.405..158R}. Along with the GPMS and HAD instruments, the probe carried a net flux radiometer for the thermal profile and heat budget \citep[NFR,][]{Sromovsky1998}; a nephelometer for cloud studies \citep[NEP,][]{1998JGR...10322891R} and an Atmospheric Structure Instrument \citep[ASI,][]{1998JGR...10322857S} to measure profiles of temperature, pressure and atmospheric density.  Measurements of the probe's transmitted radio signal (driven by an ultra-stable oscillator) allowed a reconstruction of the zonal winds with altitude \citep[Doppler Wind Experiment, DWE,][]{1998JGR...10322911A}, and attenuation of the probe-to-orbiter signal also provided information on the microwave opacity due to ammonia absorption \citep{1998JGR...10322847F}.  Comparable \textit{in situ} data for Saturn, in tandem with the wealth of remotely-sensed observations provided by Cassini, would enable a similar leap in our understanding of the solar system's second giant planet.  Finally, from the perspective of comparative planetology, improving our understanding of Saturn will provide a valuable new context for Galileo probe's measurements at Jupiter, enhancing our knowledge of this unique class of planets.

\subsection{Saturn's Dynamics and Meteorology}
\label{dynamics}

Saturn's atmosphere stands in contrast to Jupiter, with fewer large-scale vortices and a more subdued banded structure in the visible, superimposed onto hemispheric asymmetries in temperatures, cloud cover and gaseous composition as a result of Saturn's seasonal cycles (unlike Jupiter, Saturn has a considerable axial tilt of 26$^\circ$). See \cite{2009sfch.book..161W}, \cite{2009sfch.book...83F}, \cite{2009sfch.book..113D} and \cite{2009sfch.book..181N} for detailed reviews.  Despite this globally-variable atmosphere in the horizontal, a single entry probe would provide unique insights in the vertical dimension by characterising the changing environmental conditions and dynamical state as it descends from the stably-stratified middle atmosphere to the convectively-unstable troposphere.  Although \textit{in situ} probes may seem to provide one-dimensional vertical results, a horizontal dimension is also provided by reconstructing the probe trajectory during its descent, as it is buffeted by Saturn's powerful jet streams and eddies.  

\subsubsection{Atmospheric Stability and Transition Zones}
A key parameter that serves as a diagnostic of the local dynamical state of the atmosphere is the Richardson number $Ri$:

\begin{equation}
Ri = \frac{N_B^2}{\left(\pderiv{u}{z}\right)^2 + \left(\pderiv{v}{z}\right)^2} = \frac{\frac{g}{\theta}\left(\pderiv{\theta}{z}\right)}{\left(\pderiv{u}{z}\right)^2 + \left(\pderiv{v}{z}\right)^2}
\end{equation}

\noindent where $N_B$ is the frequency with which an air parcel would oscillate vertically due to buoyancy forces if perturbed from rest, also known as the \textit{Brunt V\"{a}is\"{a}l\"{a}} frequency; $\theta$ is the potential temperature and $\pderiv{\theta}{z}$ the static stability; $g$ is the gravitational acceleration and $u$ and $v$ the zonal and meridional velocities, respectively.  An entry probe can measure continuous profiles of these parameters as a function of altitude, enabling a study of stability and instability regimes as a function of depth.  Temperatures and densities in the upper atmosphere can be determined via the deceleration caused by atmospheric drag, connecting the high temperature thermosphere at nanobar pressures to the middle atmosphere at microbar and millibar pressures \citep[e.g.,][]{2004jpsm.book..185Y}. An atmospheric structure instrument would measure atmospheric pressures and temperatures throughout the descent to the clouds, and from these infer atmospheric stability and densities \citep[provided the mean molecular weight is determined by another instrument;][]{1998JGR...10322857S,2002Icar..158..410M}.  Upper atmospheric densities would be deduced from measured accelerations and from area and drag coefficients\footnote{Note that ablation sensors on the entry probe are needed to get the time-profile of Thermal Protection System (TPS) mass loss and change in area during entry.}.  The probe will sample both the radiatively-cooled upper atmosphere and also the convectively driven troposphere, precisely constraining the static stability, radiative-convective boundary (i.e., how far down does sunlight penetrate?) and the levels of the tropopause, stratopause, mesopause and homopause. Thermal structure measurements of Saturn would be directly compared to those on Jupiter to understand the energetic balance between solar heating, thermal cooling, latent heat release, wave heating and internal energy for driving the complex dynamics of all the different atmospheric layers on the giant planets, and how this balance differs as a function of distance from the Sun. 

\subsubsection{Wave Activity}
Perturbations of the temperature structure due to vertical propagation of gravity waves are expected to be common features of the stably stratified middle atmospheres either on terrestrial planets or gas giants. Wave activity is thought to be a key coupling mechanism between the convective troposphere (e.g., gravity waves and Rossby/planetary waves generated by rising plumes and vortices) and the stable middle/upper atmosphere, being responsible for transporting energy and momentum through the atmosphere and for phenomenon like the Quasi-Biennal Oscillation on Earth \citep{2001RvGeo..39..179B}, which is thought to have counterparts on Jupiter and Saturn \citep{2008Natur.453..200F}. Waves are a useful diagnostic of the background state of the atmosphere, as their propagation relies on certain critical conditions (e.g., the static stability and vertical shears on zonal winds, which cannot be revealed by remote sensing alone).  Energy and momentum transfer via waves serve as a source of both heating and cooling for the hot thermospheres, whose temperatures far exceed the expectations from solar heating alone, although the precise origins of the heating source has never been satisfactorily identified \citep[e.g.,][]{2000Icar..148..266H,2009sfch.book..181N}.  The periodicity of gravity waves measured by the Galileo probe on Jupiter permits the reconstruction of the zonal wind profile from the lower thermosphere to the upper troposphere \citep{2013GeoRL..40..472W}, permitting identification of the homopause (where molecular and eddy diffusion become comparable and gravity waves break to deposit their energy), above which the atmosphere separates into layers of different molecular species.  Understanding the propagation, periodicity and sources of wave activity on Saturn will reveal the properties of the background medium and the coupling of the ``weather layer'' to the middle atmosphere especially on how zonal and meridional circulations are forced by eddy-mean flow interactions, and facilitate direct comparison with Jupiter. 

\subsubsection{Profiling Atmospheric Winds}
\textit{In situ} exploration would tackle one of the most enduring mysteries for the giant planets - what powers and maintains the zonal winds responsible for the planetary banding, how deep do those winds penetrate into the troposphere, and what are the wind strengths in the middle atmosphere?  Remote sensing of temperature contrasts (and hence wind shears via thermal wind relationships), or inferences from the properties of atmospheric plumes at the cloud-tops \citep[e.g.,][]{2008Natur.451..437S} cannot directly address this question.  Remotely observed cloud motions are often ambiguous due to uncertainties in the cloud location; the clouds themselves may be imperfect tracers of the winds; and vertical temperature profiles (and hence wind shears) are degenerate with the atmospheric composition.  \textit{In situ} measurements of the vertical variation of winds and temperatures should resolve these ambiguities. The Galileo probe's DWE reported that jovian winds were at a minimum at the cloud tops (where most of our understanding of zonal winds and eddy-momentum fluxes originate from), and increased both above \citep{2013GeoRL..40..472W} and below \citep{1998JGR...10322911A} this level.  In the deep atmosphere, DWE demonstrated that Jupiter's winds increased to a depth of around 5 bars, and then remained roughly constant to the maximum probe depth of around 22 bars.  Similar measurements on Saturn could sample the transition region between two different circulation regimes - an upper tropospheric region where eddies cause friction to decelerate the zonal jets and air rises in cloudy zones, and a deeper tropospheric region where the circulation is reversed and eddy pumping is essential to maintain the jets and air rises in the warmer belts \citep[e.g.,][]{2009sfch.book..113D, 2011Icar..214..510F}.  A single entry probe would potentially sample both regimes,  and reconciling these two views of tropospheric circulation on Saturn would have implications for all of the giants.  Finally, direct measurements of winds in the middle atmosphere would establish the reliability of extrapolations from the jets in the cloud tops to the stratosphere in determining the general circulations of planetary stratospheres.

\subsection{Saturn's Clouds and Composition}

In Section \ref{origin} we discussed the need for reliable measurements of bulk elemental enrichments and isotopic ratios to study the formation and evolution of Saturn. Vertical profiles of atmospheric composition (both molecular and particulate) are essential to understanding the chemical, condensation and disequilibrium processes at work, in addition to the deposition of material from outside of the planet's atmosphere. The Galileo probe compositional and cloud measurements revealed an unexpectedly dry region of the jovian troposphere, depleted in clouds and volatiles \citep{1999P&SS...47.1243A}, which was consistent with ground-based observations of the probe entry into a warm cyclonic region \citep[e.g.,][]{1998JGR...10322791O}.  For this reason, the compositional profiles measured by Galileo are not thought to be globally representative of Jupiter's atmosphere, leading to a desire for multiple entry probes for different latitudes and longitudes in future missions.  Nevertheless, a single probe is essential for a more complete understanding of this class of giant planets, to enhance our knowledge of Saturn and to provide a context for improved interpretation of the Galileo probe's sampling of Jupiter's unusual meteorology.

\subsubsection{Clouds and hazes}

A poor understanding of cloud and haze formation in planetary atmospheres of our solar system may be the key parameter limiting our ability to interpret spectra of extrasolar planets and brown dwarfs \citep[e.g.,][]{2013arXiv1301.5627M}. Although equilibrium cloud condensation models \citep[ECCMs,][]{1973Icar...20..465W} combined with the sedimentation of condensates to form layers, have proven successful in explaining the broad characteristics of the planets (methane ice clouds on ice giants, ammonia ice clouds on gas giants), they remain too simplistic to reproduce the precise location, extent and microphysics of the observed cloud decks.  The Galileo probe results defied expectations of equilibrium condensation by revealing clouds bases at 0.5, 1.3 and 1.6 bar, plus tenuous structure from 2.4-3.6 bar and no evidence for a deep water cloud \citep{1999P&SS...47.1243A,2004jpsm.book...79W}.  Ammonia ice on Jupiter has only been spectroscopically identified in regions of powerful convective updrafts \citep[e.g.,][]{2002Icar..159...74B,2007Sci...318..223R}, and water ice has been detected in Voyager far-infrared spectroscopy \citep{2000Icar..145..454S}.  The spectral signature of pure ammonia ice is likely obscured by a coating or mixing with other products, such as photolytically produced hydrocarbons, hydrazine or diphosphine \citep[e.g.,][]{2010Icar..210..230S,2004jpsm.book...79W}.   The spectral properties of these mixtures are poorly known, rendering cloud remote sensing highly ambiguous.  Furthermore, Saturn's upper troposphere appears dominated by a ubiquitous haze whose composition has never been determined and is potentially unrelated to condensed volatiles (although diphosphine, P$_2$H$_4$, a product of the UV destruction of phosphine, remains an intriguing possibility).  An ECCM applied to Saturn with a $5\times$ enhancement of heavy elements over solar abundances predicts NH$_3$ condensation at 1.8 bar, NH$_4$SH near 4 bar and an aqueous ammonia cloud (merging with a water ice cloud) near 20 bar \citep{1999P&SS...47.1243A}.  However, ammonia and water ice signatures have been identified only recently, in the powerful updrafts associated with a powerful springtime storm in 2010--2011 \citep{2013Icar..226..402S}.  

The only way to resolve these questions is by \textit{in situ} sampling of the clouds and hazes formed in a planet's atmosphere, using instruments designed to measure the particle optical properties, size distributions, number and mass densities, optical depth and vertical distribution. Combined with the vertical profiles of condensable volatiles (e.g., NH$_3$, H$_2$S and H$_2$O on Saturn) and photochemically-produced species (hydrocarbons, hydrazine N$_2$H$_4$, diphosphine), this would give an estimate of the composition of Saturn's condensation clouds and upper atmospheric hazes for the first time.  Saturn's atmosphere provides the most accessible cloud decks for this study after Jupiter (condensates of NH$_3$ and H$_2$O are locked away at considerably higher pressures on the ice giants); the most useful comparison to remote sensing data (e.g., from Cassini); and the most similar composition to Jupiter for a full understanding of gas giant clouds.

\subsubsection{Atmospheric Chemistry and Mixing}

Gaseous species can be removed from the gas phase by condensation; modified by vertical mixing and photolysis; and deposited from exogenic sources (icy rings, satellites, interplanetary dust, comets, etc.), causing abundance profiles to vary with altitude and season.  Indeed, all the giant planets exhibit a rich chemistry due to the UV photolysis of key atmospheric species. Their stratospheres are dominated by the hydrocarbon products of methane photolysis \citep[e.g.,][]{2005JGRE..110.8001M}, which descend into the troposphere to be recycled by thermochemical conversion.  On Jupiter, the Galileo probe was able to measure hydrocarbon species in the 8--12 bar region, although the balance of ethane (expected to be the most abundant hydrocarbon after methane) to ethylene, propene, acetylene and propane led to suspicions that the hydrocarbon detections were instrumental rather than of atmospheric origin \citep{2009Icar..199..231W}.  Stratospheric measurements of hydrocarbons in their production region were not performed, but would be possible on Saturn with a probe. Saturn's troposphere features saturated volatiles in trace amounts above the cloud tops, but only ammonia gas is abundant enough for remote detection. H$_2$S and H$_2$O profiles above the condensation clouds have never been measured.  In addition to the volatiles, Saturn's troposphere features a host of disequilibrium species, most notably phosphine, dredged up from the deeper, warmer interior by vigorous atmospheric mixing \citep[e.g.,][]{2009Icar..202..543F}.  The abundance of PH$_3$ measured in the upper troposphere is thought to represent the abundance at its thermochemical quench level, where the vertical diffusion timescale is shorter than the thermochemical kinetics timescale.  Measurements of additional trace species in the troposphere (GeH$_4$, AsH$_3$, CO) provide constraints on the strength of atmospheric mixing from deeper, warmer levels below the clouds.  CO is of particular interest because it could be used as a probe of the deep O/H ratio of Saturn (see Section \ref{origin}). 

Detection of trace chemical species (HCN, HCP, CS, methanol, formaldehyde) and hydrogen halides \citep[HCl, HBr, HF and HI, e.g.,][]{2006Icar..185..466T,2012A&A...539A..44F} would reveal coupled chemistry due to lightning activity or shock chemistry due to planetary impacts.  In addition, the presence of oxygenated species in the upper stratosphere (CO, CO$_2$, H$_2$O) reveal the strength of exogenic influx of materials \citep[comets, interplanetary dust, e.g.,][]{1997Natur.389..159F,2010A&A...510A..88C} into the upper atmosphere of Saturn.  Sensitive mass spectrometry of these species, combined with probe measurements of atmospheric temperatures and haze properties, could reveal the processes governing the soup of atmospheric constituents on the giant planets. Once again, Saturn's trace species are expected to be the most accessible of the solar system giant after Jupiter, as volatiles and disequilibrium species (e.g., PH$_3$ and NH$_3$) have so far eluded remote 
detection on the ice giants.

\subsection{Summary of Key Atmospheric Measurements}
\label{key_atm}

A single entry probe would reveal new insights into the vertical structures of temperatures, density, chemical composition and clouds during descent through a number of different atmospheric regions, from the stable upper/middle atmosphere to the convective troposphere.  It would directly sample the condensation cloud decks and ubiquitous hazes whose composition, altitude and structure remain ambiguous due to the inherent difficulties with remote sensing.  Furthermore, it would show how Saturn's atmosphere flows at a variety of different depths above, within and below the condensate clouds. Key measurements required to address the science described in this section include:

\begin{itemize}
\item Continuous measurements of atmospheric temperature and pressure throughout the descent to study (i) stability regimes as a function of depth though transition zones (e.g., radiative-convective boundary); (ii) atmospheric drag and accelerations; and (iii) the influence of wave perturbations and cloud formation on the vertical temperature profile;
\item Determination of the vertical variation of horizontal winds using Doppler measurements of the probe's carrier frequency (driven by an ultra-stable oscillator) during the descent.  This includes a study of the depth of the zonal wind fields, as well as the first measurements of middle atmospheric winds;
\item  Vertical profiling of a host of atmospheric species via mass spectrometry, including atmospheric volatiles (water, H$_2$S and NH$_3$ in their saturated and sub-cloud regions); disequilibrium species (e.g., PH$_3$, AsH$_3$, GeH$_4$, CO) dredged from the deeper atmosphere; photochemical species (e.g., hydrocarbons and HCN in the troposphere and stratosphere; hydrazine and diphosphine in the upper troposphere) and exogenic inputs (e.g., oxygenated species in the upper atmosphere);
\item Measurements of the vertical structure and properties of Saturn's cloud and haze layers; including determinations of the particle optical properties, size distributions, number and mass densities, opacity, shapes and, potentially, their composition.
\end{itemize}

With a single entry probe, the selected entry site must be carefully studied.  Saturn's equatorial zone is one potential site for a single entry probe because of its meteorological activity that combines: the emergence of large-scale storms \citep{1991Natur.353..397S}; vertical wind shears in the troposphere \citep{2011Icar..215...62G}; upwelling enhancing volatiles and disequilbrium species \citep{2009Icar..202..543F,2011Icar..214..510F}; and a global stratospheric oscillation of the thermal field \citep{2008Natur.453..200F,2008Natur.453..196O,2011GeoRL..38.9201G}. Additionally, the strength of its equatorial eastward jet (peak velocities up to 500 m/s) poses one of the theoretical challenges to the understanding of the dynamics of fluid giant planets. Furthermore, a descent probe into Saturn's equatorial region could further constrain the influx of H$_2$O originating from the Enceladus torus \citep{2011A&A...532L...2H}.  However, it remains an open question as to how representative the equatorial region would be of Saturn's global dynamics.  Short of multiple entry probes targeted at different regions of upwelling and subsidence, near to narrow prograde jets or broader retrograde jets, a mid-latitude atmospheric region might be a more representative sample.

\section{Mission Architectures}
\label{archi}

The primary science objectives described in Sec. \ref{origin} and \ref{atm} may be addressed by an atmospheric entry probe that would descend under parachute, and start to perform {\it in situ} measurements in the stratosphere to help characterize the location and properties of the tropopause, and continue into the troposphere to pressures of at least 10 bars. All of the science objectives, except for the abundance of oxygen which may be only addressed partially, can be achieved by reaching 10 bars. Previous studies have shown that depths beyond 10 bars become increasingly more difficult to achieve for several technology reasons; for example: i) the descent time, hence the relay duration, would increase and make the relay geometry more challenging; ii) the technology for the  probe may change at pressures greater than 10 bars; iii) the opacity of the atmosphere to radio-frequencies increases with depth and may make the communication link even more challenging at higher pressures. Future studies would be needed to conduct a careful assessment of the trade-offs between science return and the added complexity of a probe that could operate at pressures greater than 10 bars. Accelerometry measurements may also be performed during the entry phase in the higher part of the stratosphere to probe the upper layers of the atmosphere prior to starting {\it in situ} measurements under parachute. 

A carrier spacecraft would be required to deliver the probe to the desired atmospheric entry point at Saturn. We have identified three possible mission configurations:

\begin{itemize}
\item {\bf Configuration 1: Probe + Carrier.} The probe would detach from the carrier spacecraft prior to probe entry. The carrier would follow the probe path and be destroyed during atmospheric entry, but may be capable of performing pre-entry science. The carrier would not be used as a radio relay to transmit the probe data to Earth. The probe would transmit its data to the ground system via a direct-to-Earth (DTE) RF link;
\item {\bf Configuration 2: Probe + Carrier/Relay.} The probe would detach from the carrier several months prior to probe entry. Subsequent to probe release, the carrier trajectory would be deflected to prepare for over-flight phasing of the probe descent location for both probe data relay as well as performing approach and flyby science;
\item {\bf Configuration 3: Probe + Orbiter.} This configuration would be similar to the Galileo Orbiter/Probe mission. The probe would detach from the orbiter several months prior to probe entry. As for Configuration 2, subsequent to probe release, the orbiter trajectory would be deflected to prepare for over-flight phasing of the probe descent location. After probe relay during over-flight, the orbiter would be placed in orbit around Saturn and continue to perform orbital science.
\end{itemize}

Configuration 1 would allow the carrier to perform months of approach science and {\it in situ} pre-entry science. In this architecture, the probe data transmission would rely solely on a Direct-to-Earth probe telecommunications link. In addition to being used as the probe relay data following completion of the probe mission, Configuration 2 would possibly also provide the capability to perform months of approach science, but in addition flyby science (for a few days). This configuration would allow many retransmissions of the probe data for redundancy. Configuration 3 would clearly be the most capable, but most costly configuration. Trade-off studies will need to be carried out to assess whether the supporting remote sensing observations may be achievable during the approach phase and a single flyby or from an orbiter.  Any of the carrier options could provide context observations but an orbiter could bring more science return in addition to supporting the probe science. The only requirement is that those data be downlinked to Earth while the spacecraft is still operating.  For example, useful observations from a Configuration 1 carrier could be made several hours before probe entry, and downlink could be accomplished in the intervening time. Finally, it may be worth studying if the emerging solar-sail propulsion technology \citep{Janhunen2014} can be considered for this option.

\subsection{Atmospheric Entry Probe}
\label{entry}

An atmospheric entry probe at Saturn would in many respects resemble the Jupiter Galileo probe. The concept was put forward for Saturn in the KRONOS mission proposal (Marty et al. 2009). Giant Planet probe concept studies have been studied by ESA in 2010\footnote{http://sci.esa.int/sre-fp/47568-pep-assessment-study-internal-final-presentation/}. As an example, the KRONOS probe had a mass of $\sim$337kg, with a 220kg deceleration module (aeroshell, thermal protection system, parachutes and separation hardware) and a 117kg descent module, including the probe structure, science instruments, and subsystems. It is anticipated that the probe architecture for this mission would be battery powered and accommodate either a DTE link or a data relay to the carrier or the orbiter.  Trades would be done to assess the complexity (and cost) of probe and telecomm link design as a function of operational depth in the atmosphere.  A representative payload for the Saturn probe that would allow addressing the science objectives identified in Sec. \ref{origin} and \ref{atm}  is shown in Table 5.

\subsection{Carrier or Orbiter}
\label{carrier}

Alternative architectures for the carrier (Configuration 1 or 2) or the orbiter (Configuration 3) would be considered, taking into account, if possible and if technologically and programmatically sound,  the heritage for outer planet/deep space missions within either ESA or NASA. As an example, the carrier or the orbiter may benefit from subsystems developed by either ESA or NASA for previous outer planet missions (for example ESA/JUICE or NASA/JUNO, or possibly NASA/ESA Cassini-Huygens). 

\subsection{Power Generation}
\label{power}

It would be worth studying whether the proposed mission architectures could be solely designed on batteries and solar power. It would require qualification of the low-intensity low-temperature (LILT) solar array cells for 9.5 AU conditions. The probe would be powered with primary batteries as were the Galileo and Huygens probes. In all three configurations, the carrier (configuration 1 and 2) or the orbiter (configuration 3) would be equipped with a combination of solar panels, secondary batteries and possibly a set of primary batteries for phases that require a high power demand, for example during the probe entry phase. Nuclear power would be considered for the carrier or the orbiter  only if available solar power technology would be found to be unfeasible.

\subsection{Interplanetary Trajectory and Entry Zone of the Probe}
\label{traj}

Many trajectory options have been identified, using both direct and gravity-assisted transfers to Saturn, and more will be identified in subsequent studies. Trajectory selection will be based on the selected carrier option, launch vehicle capabilities, and available probe thermal protection capability. The interplanetary trajectory and the probe entry location are inseparably linked.  Saturn's extensive ring system presents a severe collision hazard to an inbound probe. For various declinations of the spacecraft's approach asymptote, some latitudes will be inaccessible because the trajectories to deliver to those latitudes would impact the rings.  Also, although it is possible to adjust the inclination of the approach orbit for purposes of accessing a desired latitude, this approach can greatly increase the atmosphere-relative entry speeds, possibly driving the mission to an expensive heat shield material technology development.  During the studies, the issues of probe entry locations, approach and entry trajectories, and probe technologies must be treated together. Due to Saturn's large obliquity and the characteristics of reasonable Earth-to-Saturn transfer trajectories, the best combinations change with time. Concerning the probe entry zone, both equatorial and mid-latitude regions may be a representative location from the scientific point of view (see a discussion in Sec. \ref{key_atm}). Volatile-depleted regions are probably located at the cyclones in both poles and may also be located at the so-called ``storm-alley'' (region of low static stability able to develop updrafts and downdrafts). More generally, the peaks of westward jets can be unstable based on the stability of the wind system and eastward jets (particularly the anticyclonic branch of eastward jets) might be good locations to retrieve the deep values of volatiles at higher levels in the atmosphere \citep{2009P&SS...57.1682R}. In any case, there are several potential entry points and a dedicated study will be needed to assess their relative merit. 

\subsection{International Collaboration}
\label{coll}

In this paper, we only consider ESA/Europe and NASA/USA collaborations but collaborations with other international partners may be envisaged. One of the key probe technologies for a Saturn probe that would be new for European industry, is the heat shield material. Recent NASA studies concerning entry system performance requirements versus thermal protection system capability for a Saturn entry probe have been completed \citep{Ellerby2013}. This example is used to illustrate that, should Europe be willing to lead the probe development (as was so successfully done with Huygens), careful consideration of trade-offs would have to be made for either development of this new technology within Europe or for establishing an international collaboration with NASA. International collaboration may also be considered for other mission elements, including the carrier (and of course the orbiter if configuration 3 would be further studied), navigation, the probe data relay, the ground segment, and science payload. All three configurations would be achievable through different schemes of collaboration between Europe and the USA. As an example, configurations 1 and 2 may take the form of a combined ESA/Class-M and a NASA Discovery or New Frontiers collaboration, if such a scheme were to become programmatically feasible as it is currently not the case. Configuration 3 would be achievable through a collaboration that would involve an ESA/Class M and a NASA/Flagship mission. We do not put forward an ESA/Class L mission at this stage as the current ESA Cosmic Vision plan would not allow a new Class-L mission before the late 30's/early 40's.

\section{Characteristics of a Possible Probe Model Payload}
\label{inst}

The scientific requirements discussed above can be addressed with a suite of scientific instruments, which are given in Table~5 and discussed in the following.  Note that this list of instruments should not be considered as a unique payload complement but as guideline for some of the instruments we might wish to see on board. For example, an alternative to both the nephelometer and net flux radiometer described below are elements of the Huygens Descent Imager/Spectral Radiometer (DISR) \citep{2002SSRv..104..469T} that used the sun as a source. Ultimately, the payload of the Saturn probe would be the subject of detailed mass, power and design trades, but should seek to address the majority of the scientific goals outlined in Sec. \ref{origin} and \ref{atm}.

\subsection{Mass Spectrometry}
\label{inst_MS}

The chemical and isotopic composition of Saturn's atmosphere, and its variability, will be measured by mass spectrometry. The gas analysis systems for a Saturn Probe may benefit from a high heritage from instrumentation already flown and having provided atmospheric composition and isotope investigations. The scientific objective for the mass spectrometric investigation regarding Saturn's formation and the origin of the solar system are {\it in situ} measurements of the chemical composition and isotope abundances in the atmosphere, such as H, C, N, S, P, Ge, As, noble gases He, Ne, Ar, Kr, and Xe, and the isotopes D/H, $^{13}$C/$^{12}$C, $^{15}$N/$^{14}$N, $^{3}$He/$^{4}$He, $^{20}$Ne/$^{22}$Ne, $^{38}$Ar/$^{36}$Ar, $^{36}$Ar/$^{40}$Ar, and those of Kr and Xe. 

At Jupiter, the Galileo Probe Mass Spectrometer (GPMS) experiment \citep{1992SSRv...60..111N} was used, which was designed to measure the chemical and isotopic composition of Jupiter's atmosphere in the pressure range from 0.15 to 20 bar by {\it in situ} sampling of the ambient atmospheric gas. The GPMS consisted of a  gas sampling system that was connected to a quadrupole mass spectrometer. The gas sampling system also had two sample enrichment cells, allowing for enrichments of hydrocarbons by a factor 100 to 500, and one noble gas analysis cell with an enrichment factor of about 10. Low abundance noble gases can be measured by using an enrichment cell, as used on the Galileo mission \citep{1996Sci...272..846N}. From GPMS measurements the Jupiter He/H$_2$ ratio was determined as 0.156 $\pm$ 0.006. To improve the accuracy of the measurement of the He/H$_2$ ratio and isotopic ratios by mass spectrometry the use of reference gases will be necessary. On the Rosetta mission the ROSINA experiment carries for each mass spectrometer a gas calibration unit \citep{2007SSRv..128..745B}. Similarly, the SAM experiment on the Curiosity rover can use either a gas sample from its on-board calibration cell or utilize one of the six individual metal calibration cups on the sample manipulation system \citep{2012SSRv..170..401M}.

A major consideration for the mass spectrometric analysis is how to distinguish between different molecular species with the same nominal mass, e.g. N$_2$ and CO, from the complex mixture of gases in Saturn's atmosphere. There are two approaches available, one is high resolution mass spectrometry with sufficient mass resolution to resolve the isobaric interferences, and the other is chemical pre-separation of the sample followed by low resolution mass spectrometry. 

\subsubsection{High Resolution Mass Spectrometry}
\label{inst_HRMS}

Probably the first composition experiment with high resolution mass spectrometry is the  ROSINA experiment on the Rosetta mission \citep{2007SSRv..128..745B} which has a Reflectron-Time-of-Flight (RTOF) instrument with a mass resolution of about $m/\Delta m$ = 5,000 at 50\% peak height \citep{2006IJMSp.251...73S}, Double-Focussing Mass Spectrometer (DFMS) with a mass resolution of about $m/\Delta m$ = 9,000 at 50\% peak height, and a pressure gauge. Determination of isotope ratios at the 1\% accuracy level has been accomplished during the cruise phase. A time-of-flight instrument with even more mass resolution has been developed for possible application in Titan's atmosphere, which uses a multi-pass time-of flight configuration \citep{waite2012}. Typical mass resolutions are $m/\Delta m$ = 13,500 at 50\% peak height and 8,500 at 20\% peak height. In bunch mode the mass resolution can be increased to 59,000 (at 50\% peak height). Again, determination of isotope ratios at the 1\% accuracy level has been accomplished. An alternative multi-pass time-of-flight instrument has been developed by \cite{Okumura2004}, which uses electric sectors instead of ion mirrors for time and space focussing. Mass resolutions up to $m/\Delta m$ = 350,000 have been reported \citep{Toyoda2003}.

Recently, a new type of mass spectrometer, the Orbitrap mass spectrometer, was introduced \citep{Makarov2000,Hu2005}, which uses ion confinement in a harmonic electrostatic potential. The Orbitrap mass spectrometer is a Fourier-Transform type mass spectrometer, and it allows for very high mass resolutions in a compact package. For example, using an Orbitrap mass spectrometer for laboratory studies of chemical processes in Titan's atmosphere, mass resolutions of $m/\Delta m$ = 100,000 have been accomplished up to m/z = 400 \citep{Hoerst2012}, and $m/\Delta m$ = 190,000 at 50\% peak height and m/z = 56 in a prototype instrument for the JUICE mission \citep{2013LPI....44.2888B}. 

\subsubsection{Low Resolution Mass Spectrometry with Chemical Pre-Processing}
\label{inst_LRMS}

The alternative approach to high resolution mass spectrometry, which was used successfully on many missions so far, is to use a simpler low resolution mass spectrometer together with a chemical processing of the sample to separate or eliminate isobaric interferences. One established way is to use chromatographic columns with dedicated chemical specificity for a separation of chemical substances before mass spectrometric analysis. The Gas-Chromatograph Mass Spectrometer (GCMS) of the Huygens Probe is a good example of such an instrument \citep{Niemann2002,2005Natur.438..779N,2010JGRE..11512006N}. The Huygens Probe GCMS has three chromatographic columns, one column for separation of CO and N$_2$ and other stable gases, the second column for separation of nitriles and other organics with up to three carbon atoms, and the third column for the separation of C$_3$ through C$_8$ saturated and unsaturated hydrocarbons and nitriles of up to C$_4$. The GCMS was also equipped with a chemical scrubber cell for noble gas analysis and a sample enrichment cell for selective measurement of high boiling point carbon containing constituents.  A quadrupole mass spectrometer was used for mass analysis with a mass range from 2 to 141 amu, which is able to measure isotope ratios with an accuracy of ~1\%. Newer examples of GCMS instrumentation are the Ptolemy instrument on the Rosetta lander for the measurement of stable isotopes of key elements \citep{2007SSRv..128..363W}, which uses an ion trap mass spectrometer, the COSAC instrument also on the Rosetta lander for the characterization of surface and sub-surface samples \citep{2007SSRv..128..257G}, which uses a time-of-flight mass spectrometer, and the SAM experiment on the Curiosity rover \citep{2012SSRv..170..401M}, which uses a classical quadrupole mass spectrometer.

\subsubsection{Summary of Mass Spectrometry}
\label{sum_MS}
So far in most missions the chemical pre-separation was the technique used to avoid isobaric interferences in the mass spectra, with the exception of the mass spectrometer experiment ROSINA on the Rosetta orbiter. Chemical pre-separation works well, but by choosing chromatographic columns with a certain chemical specificity one makes a pre-selection of the species to be investigated in detail. This possibly is a limitation when exploring an object where little is known. Also, gas chromatographic systems with several columns are rather complex systems, both to build and to operate (see the SAM instrument as a state-of-the art example of this technique \citep{2012SSRv..170..401M}).

In recent years there has been a significant development of compact mass spectrometers that offer high mass resolution, as outlined above, and  these developments are still ongoing. Thus, solving the problem of isobaric interferences in the mass spectra by mass resolution can be addressed by mass spectrometry alone and one should seriously consider using high resolution mass spectrometry for a future mission to probe Saturn's atmosphere. After all, no {\it a priori} knowledge of the chemical composition has to be assumed. In addition, with modern time-of-flight mass spectrometers mass ranges beyond 1000 amu are not a problem at all, which would have been useful to investigate Titan's atmosphere. Nevertheless, some chemical pre-selection may still be considered even for high resolution mass spectrometry. For example, the cryotrapping technique, which has a long history in the laboratory, would enable detection of noble gases at abundances as low as 0.02 ppb \citep{waite2012}.

\subsubsection{Tunable Laser System }
\label{inst_TLS}
 
A Tunable Laser Spectrometer (TLS) \citep{Durry2002} can be employed as part of a GC system to measure the isotopic ratios to a high accuracy of specific molecules, e.g. H$_2$O, NH$_3$, CH$_4$, CO$_2$ and others. TLS employs ultra-high spectral resolution (0.0005~cm$^{-1}$) tunable laser absorption spectroscopy in the near infra-red (IR) to mid-IR spectral region. TLS  is a direct non-invasive, simple technique that for small mass and volume can produce remarkable sensitivities at the sub-ppb level for gas detection. Species abundances can be measured with accuracies of a few \%, and isotope determinations  are with about 0.1\% accuracy. With a TLS system one can derive the isotopic ratios of D/H, $^{18}$O/$^{16}$O, $^{13}$C/$^{12}$C, $^{18}$O/$^{16}$O, and $^{17}$O/$^{16}$O. 

For example, TLS was developed for application in the Mars atmosphere \citep{2004cosp...35.2115L}, within the ExoMars mission; a recent implementation of a TLS system was for the Phobos Grunt mission \citep{2010ApPhB..99..339D}, and is on the  SAM instrument \citep{2011P&SS...59..271W}, which was used to measure the  isotopic ratios of D/H and of $^{18}$O/$^{16}$O in water and $^{13}$C/$^{12}$C, $^{18}$O/$^{16}$O, $^{17}$O/$^{16}$O, and $^{13}$C$^{18}$O/$^{12}$C$^{16}$O in carbon dioxide in the Martian atmosphere \citep{2013Sci...341..260W}.

\subsection{Helium Abundance Detector}
\label{inst_HAD}

The Helium Abundance Detector (HAD), as it was used on the Galileo mission \citep{von Zahn1992}, basically measures the refractive index of the atmosphere in the pressure range of 2--10 bar. The refractive index is a function of the composition of the sampled gas, and since the jovian atmosphere consists of mostly of H$_2$ and He, to more than 99.5\%, the refractive index is a direct measure of the He/H$_2$ ratio.  The refractive index can be measured by any two-beam interferometer, where one beam passes through a reference gas and the other beam through atmospheric gas. The difference in the optical path gives the difference in refractive index between the reference and atmospheric gas. For the Galileo mission a Jamin-Mascart interferometer was used, because of its simple and compact design, with an expected accuracy of the He/H$_2$  ratio of $\pm$ 0.0015. The accomplished measurement of the He mole fraction gave  0.1350 $\pm$ 0.0027 \citep{1998JGR...10322815V}, with a somewhat lower  accuracy when expected, but still better than is possible by a mass spectrometric measurement.

\subsection{Atmospheric Structure Instrument}
\label{inst_ASI}

The key {\it in situ} measurements by an Atmospheric Structure Instrument (ASI) would be the accelerometry during the probe entry phase and pressure, temperature and mean molecular weight during descent. The atmospheric density is derived from these measurements. There is strong heritage from the Huygens ASI experiment (HASI) of the Cassini-Huygens mission \citep{Fulchignoni2002}. Furthermore, these types of sensors have been selected for NASA's Mars Science Laboratory (MSL) and are part of the meteorological package of ESA's Exomars mission. {\it In situ} atmospheric structure measurements are essential for the investigation of the planet's atmospheric structure and dynamics. The estimation of the temperature lapse rate can be used to identify the presence of condensation and possible clouds, to distinguish between saturated and unsaturated, stable and conditionally stable regions. The variations in the density, pressure and temperature profiles provide information on the atmospheric stability and stratification, on the presence of winds, thermal tides, waves and turbulence in the atmosphere. A typical Atmospheric Structure Instrument would consist of three primary sensor packages: a three-axis accelerometer, a pressure profile instrument  and temperature sensors. It would start to operate right at the beginning of the entry phase of the probe, sensing the atmospheric drag experienced during entry. Direct pressure and temperature measurement could be performed by the sensors having access to the atmospheric flow from the earliest portion of the descent until the end of the probe mission at approximately 10 bar.

ASI data will also contribute to the analysis of the atmospheric composition, since it will monitor the acceleration experienced by the probe during the whole descent phase. ASI  will provide the unique direct measurements of pressure and temperature through sensors having access to the atmospheric flow.

\subsubsection{Accelerometers}
 The accelerator package, a 3-axis accelerometer, should be placed as close as possible to the center of mass of the entry probe. Like on Huygens, the main sensor could be a highly sensitive servo accelerometer aligned along the vertical axis of the Probe, with a resolution of 1 to 10$\times$10$^{-5}$ m s$^{-2}$  with an accuracy of 1\%. Probe acceleration can be measured in the range of 0--2000 m s$^{-2}$  \citep{Zarnecki2004}. Assuming the HASI accelerometer performance at Titan, a noise level of about 3$\times$10$^{-8}$ m s$^{-2}$ is expected. The exact performance achievable, in terms of the accuracy of the derived atmospheric density, will also depend on the probe ballistic coefficients, entry speed and drag coefficient.

\subsubsection{Temperature sensors}
As in the Huygens Probe, the temperature sensors will use platinum resistance thermometers. These are exposed to the atmospheric flow and are effectively thermally isolated from the support structure. The principle of measurement is based on the variation of the resistance of the metallic wire with temperature. TEM has been designed to have a good thermal coupling between the sensor and the atmosphere and to achieve high accuracy and resolution. Over the temperature range of 60--330~K these sensors maintain an accuracy of 0.1~K with a resolution of 0.02~K. 

\subsubsection{Pressure Profile Instrument} 
The Pressure Profile Instrument (PTI) will measure the pressure during the entire descent with an accuracy of 1\% and a resolution of 10$^{-6}$ bar. The atmospheric flow is conveyed through a Kiel probe inside the probe where the transducers and related electronic are located. The transducers are silicon capacitive sensors with pressure dependant dielectricum. The pressure sensor contains as dielectricum a small vacuum chamber between the two electrode plates, with the external pressure defining the distance of these plates. Detectors with diaphragms of different pressure sensitivity will be utilized to cover the pressure range to $\sim$10 bar. The pressure is derived as a frequency measurement (within 3--20~kHz range) and the measurements internally compensate for thermal and radiation influences.

\subsection{Doppler Wind Experiment}
\label{inst_DWE}
The primary goal of a Doppler Wind Experiment (DWE) on a Saturn probe would be to measure a vertical profile of the zonal (east-west) winds along the probe descent path. A secondary goal of the DWE is to detect, characterize, and quantify microstructure in the probe descent dynamics, including probe spin, swing, aerodynamic buffeting and atmospheric turbulence, and to detect regions of wind shear and atmospheric wave phenomena. Because of the need for accurate probe and carrier trajectories for making the Doppler wind measurement, the DWE must be closely coordinated with the navigation and radiometric tracking of the carrier, and the probe entry and descent trajectory reconstructions. A Doppler Wind Experiment could be designed to work with a probe DTE communication architecture or a probe-to-relay architecture.  Both options include ultra-stable oscillator (USO) requirements and differ only in the angle of entry and DTE geometry requirements.  For relay, the system comprises a probe and a carrier USO as part of the probe-carrier communication package.  The experiment would benefit from the heritage of both the Galileo and Huygens Doppler Wind Experiments \citep{1998JGR...10322911A, Bird2002}.

\subsection{Nephelometer}
\label{inst_NEF}

The composition and precise location of cloud layers in Saturn are largely unknown. They may be composed of ammonia, ammonium hydrosulfide, or simply water. Because of this relative paucity of information on Saturn's clouds, the demands we place on a cloud particle sensor, a nephelometer, are significant. Such an instrument would have little heritage in planetary exploration, beyond the one on the Galileo probe. There are similar laser driven, fiber fed nephelometers used in very similar settings on Earth \citep[e.g.,][]{Barkey2001, Barkey1999, Gayet1997}. However, these were shrouded designs, which is mass-prohibitive on a planetary probe, and they also did not attempt to measure the polarization ratio phase function, because they knew their aerosols were water.  The polarization modulation technique that we are proposing was first described by \cite{Hunt1973}, and has been used in laboratory settings by several groups \citep[e.g.,][]{Kuik1991}. While the precise implementation of the instrument is novel to planetary science, and the polarization modulation technique is also new to an {\it in situ} instrument, the technology needed to carry out this instrument is all relatively modest. This nephelometer would measure not only the amplitude phase function of the light scattered by the clouds from a laser source on the probe, but also the polarization ratio phase function as well.  

\subsection{Net Energy Flux Radiometer}
\label{inst_NFR}
A Net Energy Flux Radiometer (NFR) measures the thermal profile and heat budget in the atmosphere. Such a NFR instrument was part of the scientific payload of the Galileo mission \citep{Sromovsky1992}, which measured the vertical profile of upward and downward radiation fluxes in the region between 0.44 to 14 bar region \citep{Sromovsky1998}.  Radiation was measured in five wavelength bands, 0.3--3.5 $\mu$m (total solar radiation), 0.6--3.5 $\mu$m (total solar radiation in methane absorption region), 3--500 $ \mu$m (deposition and loss of thermal radiation), 3.5--5.8 $\mu$m (water vapor and cloud structure), and 14--35 $\mu$m (water vapor). On Galileo, NFR found signatures of NH$_3$ ice clouds and NH$_4$SH clouds \citep{Sromovsky1998}, however, water fraction was found to be much lower than solar and no water clouds could be indentified.

\section{Conclusions}
\label{cls}

In this paper, we have shown that the {\it in situ} exploration of Saturn can address two major science themes: the formation history of our solar system and the processes at work in the atmospheres of giant planets. We provided a list of recommended measurements in Saturn's atmosphere that would allow disentangling between the existing giant planets formation scenarios and the different volatile reservoirs from which the solar system bodies were assembled. Moreover, we illustrated how an entry probe would reveal new insights concerning the vertical structures of temperatures, density, chemical composition and clouds during atmospheric descent. In this context, the top level science goals of a Saturn probe mission would be the determination of:

\begin{enumerate}

\item the atmospheric temperature, pressure and mean molecular weight profiles;
\item the abundances of cosmogenically abundant species C, N, S and O;
\item the abundances of chemically inert noble gases He, Ne, Xe, Kr and Ar;
\item the isotopic ratios in hydrogen, oxygen, carbon, nitrogen, He, Ne, Xe, Kr and Ar;
\item the abundances of minor species delivered by vertical mixing (e.g., P, As, Ge) from the deeper troposphere, photochemical species (e.g., hydrocarbons, HCN, hydrazine and diphosphine) in the troposphere and exogenic inputs (oxygenated species) in the upper atmosphere;
\item the particle optical properties, size distributions, number and mass densities, opacity, shapes and composition.

\end{enumerate}

\noindent Additional {\it in situ} science measurements aiming at investigating the global electric circuit on Saturn could be also considered (measurement of the Schumann resonances, determination of the vertical profile of conductivity and the spectral power of Saturn lightning at frequencies below the ionospheric cutoff, etc).

We advocated that a Saturn mission incorporating elements of {\it in situ} exploration should form an essential element of ESA and NASA's future cornerstone missions. We described the concept of a Saturn probe as the next natural step beyond Galileo's {\it in situ} exploration of Jupiter, and the Cassini spacecraft's orbital reconnaissance of Saturn. Several missions designs have been discussed, all including a spacecraft carrier/orbiter and a probe that would derive from the KRONOS concept previously proposed to ESA \citep{2009ExA....23..947M}. International collaborations, in particular between NASA/USA and ESA/Europe may be envisaged in the future to enable the success of a mission devoted to the {\it in situ} exploration of Saturn.

\section*{Acknowledgements}

O.M. acknowledges support from CNES. L.N.F was supported by a Royal Society Research Fellowship at the University of Oxford. P.W.\ acknowledges support from the Swiss National Science Foundation. O. V. acknowledges support from the KU Leuven IDO project IDO/10/2013 and from the FWO Postdoctoral Fellowship program. A. S. L and R. H. were supported by the Spanish MICIIN project AYA2012-36666 with FEDER support, PRICIT-S2009/ESP-1496, Grupos Gobierno Vasco IT765-13 and UPV/EHU UFI11/55.

\clearpage

\begin{table*}[h]
\begin{center}
\caption[]{Compositions of the atmospheres of Jupiter and Saturn (major volatiles)}
\rotatebox{90}{\scriptsize{\begin{tabular}{lcccccc}
\hline
\noalign{\smallskip}
				& \multicolumn{3}{c}{Jupiter}													& \multicolumn{3}{c}{Saturn}												\\
Species			&X/H$_2$					& $\Delta$(X/H$_2$)	& Reference					& X/H$_2$				& $\Delta$(X/H$_2$)		& Reference				\\
\hline
CH$_4$			& $2.37 \times 10^{-03}$		& 5.70 $\times 10^{-04}$	& \cite{2004Icar..171..153W}		& 5.33 $\times 10^{-3}$		& 0.23 $\times 10^{-3}$		& \cite{2009Icar..199..351F}	\\		
NH$_3$			& $6.64 \times 10^{-04}$		& 2.54 $\times 10^{-04}$	& \cite{2004Icar..171..153W}		& 1.04--5.78 $\times 10^{-4}$ 	& --						& \cite{2011Icar..214..510F}	\\
H$_2$O$^{\rm(a)}$	& $4.90 \times 10^{-04}$		& 1.60 $\times 10^{-04}$	& \cite{2004Icar..171..153W}		& 2.0 $\times 10^{-07}$		& --						& \cite{1997AA...321L..13D}	\\		
PH$_3$			& $2.11 \times 10^{-06}$		& 1.00 $\times 10^{-07}$	& \cite{2009Icar..202..543F}		& 7.30 $\times 10^{-06}$		& 0.48 $\times 10^{-06}$		& \cite{2009Icar..202..543F}	\\		
H$_2$S			& $8.90 \times 10^{-05}$		& 2.10 $\times 10^{-05}$	& \cite{2004Icar..171..153W}		& 3.90 $\times 10^{-04}$		& --						& \cite{1989Icar...80...77B}	\\		
He				& $1.36 \times 10^{-01}$		& 2.70 $\times 10^{-03}$	& \cite{1998JGR...10322815V}		& 1.35 $\times 10^{-1}$		& 0.25 $\times 10^{-1}$		& \cite{2000Icar..144..124C}	\\		
Ne$^{\rm(b)}$		& $3.0 \times 10^{-05}$		& --					& \cite{1998JGR...10322831N}		& --						& --						&						\\		
Ar				& $1.82 \times 10^{-05}$		& 3.60 $\times 10^{-06}$	& \cite{2000JGR...10515061M}		& --						& --						& 						\\		
Kr				& $9.30 \times 10^{-09}$		& 1.70 $\times 10^{-09}$	& \cite{2000JGR...10515061M}		& --						& --						& 						\\		
Xe				& $8.90 \times 10^{-10}$		& 1.70 $\times 10^{-10}$	& \cite{2000JGR...10515061M}		& --						& --						& 						\\				
\hline
\end{tabular}}}\\
$^{\rm(a)}$This is a lower limit; $^{\rm(b)}$this is an upper limit.
\end{center}
\label{table1}
\end{table*}

\clearpage

\begin{table}
\begin{center}
\caption[]{Isotopic ratios in Jupiter and Saturn}
\rotatebox{90}{\scriptsize{\begin{tabular}{lcccccc}
\hline
\noalign{\smallskip}
							& \multicolumn{3}{c}{Jupiter}														& \multicolumn{3}{c}{Saturn}										\\
Isotopic ratio					& $\eta$					& $\Delta\eta$				& Reference					& $\eta$					& $\Delta\eta$							& Reference			\\
\hline
D/H (in H$_2$)					& 2.60 $\times$ 10$^{-5}$		& 0.70 $\times$ 10$^{-5}$		& \cite{1998JGR...10322831N}		& 1.70 $\times$ 10$^{-5}$		& $^{+0.75}_{-0.45}$ $\times$ 10$^{-05}$	& \cite{2001AA...370..610L}	\\
$^3$He/$^4$He				& 1.66 $\times$ 10$^{-4}$		& 0.05 $\times$ 10$^{-4}$		& \cite{1998JGR...10322831N}		& --						& --									&  					\\
$^{12}$C/$^{13}$C (in CH$_4$)	& 92.6					& $^{+4.5}_{-4.1}$			& \cite{1996Sci...272..846N}		& 91.8					& $^{+8.4}_{-7.8}$ 						& \cite{2009Icar..199..351F}	\\
$^{14}$N/$^{15}$N (in NH$_3$)	& 434.8					& $^{+65}_{-50}$			& \cite{2004Icar..171..153W}		& --						& --									&  					\\
$^{20}$Ne/$^{22}$Ne			& 13.0					& 2.0						& \cite{2000JGR...10515061M}		& --						& --									&  					\\
$^{36}$Ar/$^{38}$Ar				& 5.6						& 0.25					& \cite{2000JGR...10515061M}		& --						& --									&  					\\
$^{128}$Xe/total Xe				& 0.018					& 0.002					& \cite{2003PSS...51..105A}		& --						& --									&  					\\
$^{129}$Xe/total Xe				& 0.285					& 0.021					& \cite{2003PSS...51..105A}		& --						& --									&  					\\
$^{130}$Xe/total Xe				& 0.038					& 0.005					& \cite{2003PSS...51..105A}		& --						& --									&  					\\
$^{131}$Xe/total Xe				& 0.203					& 0.018					& \cite{2003PSS...51..105A}		& --						& --									&  					\\
$^{132}$Xe/total Xe				& 0.290					& 0.020					& \cite{2003PSS...51..105A}		& --						& --									&  					\\
$^{134}$Xe/total Xe				& 0.091					& 0.007					& \cite{2003PSS...51..105A}		& --						& --									&  					\\
$^{136}$Xe/total Xe				& 0.076					& 0.009					& \cite{2003PSS...51..105A}		& --						& --									&  					\\
\hline
\end{tabular}}}
\end{center}
\label{table2}
\end{table}

\clearpage

\begin{table}[h]
\begin{center}
\caption[]{Enrichments in Jupiter and Saturn relatives to Protosun}
\small{\begin{tabular}{lcccccc}
\hline
\noalign{\smallskip}
				& \multicolumn{2}{c}{Jupiter}				& \multicolumn{2}{c}{Saturn}				\\
Species			& E			& $\Delta$E$^{\rm(a)}$		& E				& $\Delta$E$^{\rm(a)}$	\\
\hline
C				& 4.40		& 1.14					& 9.90			& 1.05				\\		
N				& 4.18		& 2.08					& 0.53--4.07		& --					\\
O$^{\rm(b)}$		& 0.42		& 0.15					& $\sim$10$^{-4}$	& --		\\		
P				& 3.34		& 0.36					& 11.54			& 1.35				\\		
S				& 2.94		& 0.70					& 15.87			& --					\\		
He				& 0.72		& 0.04					& 0.71			& 0.14				\\		
Ne$^{\rm(c)}$		& 0.12		& --						& --				& --					\\		
Ar				& 2.62		& 0.86					& --				& --					\\		
Kr				& 2.23		& 0.61					& --				& --					\\		
Xe				& 2.18		& 0.61					& --				& --					\\		
		
\hline
\end{tabular}}\\
$^{\rm(a)}$Error is defined as ($\Delta$E/E)$^2$ =  ($\Delta$X/X$_{\rm planet}$)$^2$ + ($\Delta$X/X$_{\rm Protosun}$)$^2$; $^{\rm(b)}$this is a lower limit; $^{\rm(c)}$this is an upper limit.
\end{center}
\label{table3}
\end{table}

\clearpage

\begin{table*}[h]
\begin{center}
\caption[]{Elemental abundances in the Sun and Protosun}
\small{\begin{tabular}{lccccc}
\hline
\noalign{\smallskip}
Element	&	Solar dex		&	Protosolar dex		&	$\Delta$dex	& Protosolar X/H$_2$		& 	$\Delta$(X/H$_2$)	\\
\hline 
\noalign{\smallskip}
C		&	8.39			&	8.43				&	0.04			& 	5.38 $\times$ 10$^{-04}$	& 	5.19 $\times$ 10$^{-05}$ 	\\
N		&	7.86			&      7.90				&     	0.12			&	1.59 $\times$ 10$^{-04}$	& 	5.06$\times$ 10$^{-05}$ 	\\
O		&	8.73			&	8.77				&	0.07			&	1.18 $\times$ 10$^{-03}$	&	2.06$\times$ 10$^{-04}$ 	\\
P		&	5.46			&	5.50				&	0.04			&	6.32$\times$ 10$^{-07}$	&	6.10$\times$ 10$^{-08}$  	\\
S		&	7.14			&	7.18				&	0.01			&	3.03$\times$ 10$^{-05}$	&	7.05$\times$ 10$^{-07}$ 	\\
He		&	10.93		&	10.98			&	0.02			&	1.89$\times$ 10$^{-01}$	&	8.90$\times$ 10$^{-03}$ 	\\
Ne		&	8.05			&	8.09				&	0.10			&	2.46$\times$ 10$^{-04}$	&	6.37$\times$ 10$^{-05}$ 	\\
Ar		&	6.50			&	6.54				&	0.10			&	6.93$\times$ 10$^{-06}$	&	1.80$\times$ 10$^{-06}$ 	\\
Kr		&	3.28			&	3.32				&	0.08			&	4.18$\times$ 10$^{-09}$	&	8.45$\times$ 10$^{-10}$ 	\\
Xe		&	2.27			&	2.31				&	0.08			&	4.08$\times$ 10$^{-10}$	&	8.26$\times$ 10$^{-11}$ 	\\
\hline
\end{tabular}}\\
Data from \cite{2009LanB...4B...44L} with values of corrections for protosolar abundances (+0.05 dex (He) and +0.04 dex (others)) taken from \cite{2009ARA&A..47..481A}.
\end{center}
\label{table4}
\end{table*}

\clearpage

\begin{table*}[h]
\begin{center}
\caption[]{Measurement requirements}
\small{\begin{tabular}{ll}
\hline
\noalign{\smallskip}
Instrument					& Measurement									\\
\hline
Mass spectrometer				& Elemental and chemical composition					\\
							& Isotopic composition								\\
							& High molecular mass organics						\\
Helium abundance detector		& Accurate He/H$_2$ ratio							\\
Atmospheric Structure Instrument	& Pressure, temperature, density, molecular weight profile	\\
Doppler Wind Experiment			& Measure winds, speed and direction					\\							
Nephelometer					& Cloud structure									\\
							& Solid/liquid particles								\\
Net-flux radiometer				& Thermal/solar energy								\\									
\hline
\end{tabular}}
\end{center}
\label{table5}
\end{table*}

\clearpage

\clearpage

\clearpage

\end{document}